\documentclass{aa}  
\usepackage{answers}
\usepackage{amsmath}
\usepackage{amssymb}
\usepackage{threeparttable}

\usepackage{graphicx}
\usepackage{txfonts}
\usepackage{natbib,twoopt}
\usepackage[breaklinks=true]{hyperref} 
\bibpunct{(}{)}{;}{a}{}{,} 
\makeatletter
\newcommandtwoopt{\citeads}[3][][]{\href{http://adsabs.harvard.edu/abs/#3}%
{\def\hyper@linkstart##1##2{}%
\let\hyper@linkend\@empty\citealp[#1][#2]{#3}}}
\newcommandtwoopt{\citepads}[3][][]{\href{http://adsabs.harvard.edu/abs/#3}%
{\def\hyper@linkstart##1##2{}%
\let\hyper@linkend\@empty\citep[#1][#2]{#3}}}
\newcommandtwoopt{\citetads}[3][][]{\href{http://adsabs.harvard.edu/abs/#3}%
{\def\hyper@linkstart##1##2{}%
\let\hyper@linkend\@empty\citet[#1][#2]{#3}}}
\newcommandtwoopt{\citeyearads}[3][][]%
{\href{http://adsabs.harvard.edu/abs/#3}
{\def\hyper@linkstart##1##2{}%
\let\hyper@linkend\@empty\citeyear[#1][#2]{#3}}}
\makeatother

\begin{document}

   \title{Effect of dust on Kelvin-Helmholtz instabilities}

   \subtitle{}

   \author{T. Hendrix
          \and
          R. Keppens
          }

   \institute{Centre for mathematical Plasma Astrophysics, Department of Mathematics, KU Leuven,\\ Celestijnenlaan 200B, 3001 Leuven, Belgium \\
              \email{tom.hendrix@wis.kuleuven.be}}

   \date{}

 
  \abstract
   {Dust is present in a large variety of astrophysical fluids, ranging from tori around supermassive black holes to molecular clouds, protoplanetary discs, and cometary outflows. In many such fluids, shearing flows are present, which can lead to the formation of Kelvin-Helmholtz instabilities (KHI) and may change the properties and structures of the fluid through processes such as mixing and clumping of dust. }
   {We study the effects of dust on the KHI by performing numerical hydrodynamical dust+gas simulations. We investigate how the presence of dust changes the growth rates of the KHI in 2D and 3D and how the KHI redistributes and clumps dust. We investigate if similarities can be found between the structures in 3D KHI and those seen in observations of molecular clouds.}
   {We perform numerical multifluid hydrodynamical simulations with in addition to the gas a number of dust fluids. Each dust fluid represents a portion of the particle size-distribution. We study how dust-to-gas mass density ratios between 0.01 and 1 alter the growth rate in the linear phase of the KHI. We do this for a wide range of perturbation wavelengths, and compare these values to the analytical gas-only growth rates. As the formation of high-density dust structures is of interest in many astrophysical environments, we scale our simulations with physical quantities that are similar to values in molecular clouds.}
   {Large differences in dynamics are seen for different grain sizes. We demonstrate that high dust-to-gas ratios significantly reduce the growth rate of the KHI, especially for short wavelengths.  We compare the dynamics in 2D and 3D simulations, where the latter demonstrates additional full 3D instabilities during the non-linear phase, leading to increased dust densities. We compare the structures formed by the KHI in 3D simulations with those in molecular clouds and see how the column density distribution of the simulation shares similarities with log-normal distributions with power-law tails sometimes seen in observations of molecular clouds. }
   {}

   \keywords{Instabilities --
                Hydrodynamics --
                ISM: clouds --
                ISM: kinematics and dynamics --
                dust
               }

   \maketitle
%

\section{Introduction}

The Kelvin-Helmholtz instability (KHI) occurs at the interface of shearing fluids and taps energy from the velocity difference to create rotating structures in-between. It has been studied repeatedly, on the one hand, because it has important effects on the surrounding environment, such as mixing and redistribution of matter and energy on various scales; on the other hand, it is of relevance in many different branches of fluid studies, such as oceanic circulation \citep{Haren}, winds on planet surfaces, magnetic reconnection in the solar corona \citep{2003SoPh..214..107L}, interaction between comet tails and the solar wind \citep{1980SSRv...25....3E}, astrophysical jets \citep{2006A&A...447....9B}, and many others. In many fluids, however, the fluid does not consist of only one smooth mixture, but one can think of the flow as loaded with extra ``impurities", having different physical properties as the particles making up the flow itself. An example of such impurities are dust grains. Even though the number of dust particles per volume is typically low compared to that of the gas because of the large size and mass of the dust particles compared to typical fluid particles, they can still be of importance on the dynamics of the system. Furthermore, the dynamics of the dust itself can be of major interest, as for example in protoplanetary disks where the initially dilute dust particles embedded in the gas can separate itself from the gas and eventually clump to form planetesimals and later planets. However, how the dust can actually grow to large macroscopic sizes is not fully understood. To explain the dust growth in protoplanetary disks, high dust densities are required, which can be obtained through (the interplay of) several instabilities. Often an instability can dynamically raise the local dust density to values high enough to initiate a gravitational collapse. \citet{2012A&A...545A.134M} used numerical simulations to show that the Rossby wave instability in disks can trap dust. \citet{2012MNRAS.423..437J} discuss how dust in disks is influenced by the magnetorotational instability. In protoplanetary disks, the KHI can be of major importance. As the dust, which moves closer to the Keplerian velocity than the gas, settles gravitationally towards the midplane, it can accelerate the gas in the midplane to higher velocities, potentially causing a shear instability between the faster moving gas in the midplane and the slower gas above and below. In this way, the KHI can potentially disrupt the growths to higher density in the midplane, or dynamically cause higher densities in some regions. This has been studied analytically as well as numerically in  \citet{2006ApJ...641.1131M}, \citet{2006ApJ...643.1219J}, and \citet{2009ApJ...691..907B}. \\

In systems where dust is typically found, particles have a large (and often unknown) variety of sizes and compositions; therefore, including the effect of dust in the analysis of the dynamics of these systems causes a significant increase in numerical and analytical complexity. Furthermore, the density of the dust is several orders of magnitude lower than the local gas density in most cases, and as a consequence, the contribution of dust on the dynamics has often been ignored. In the last few years, several works have been presented with different numerical approaches towards the incorporation of dust. Examples range from fully particle-based two-fluid smoothed-particle hydrodynamics (SPH) methods \citep{2011MNRAS.415.3319C,2012MNRAS.420.2345L} to hybrid particle-fluid methods \citep{2006ApJ...643.1219J,2010JCoPh.229.3916M} and two-fluid hydrodynamic models \citep{2006A&A...453.1129P,2011ApJ...734L..26V,2012A&A...545A.134M}.\\

Here, we gain insight in the consequence of having dust particles embedded in a fluid by investigating the effect of dust on the KHI. We do this by making 2D and 3D simulations of the dusty KHI with the fluid code MPI-AMRVAC \citep{2012JCoPh.231..718K}. While many previous studies that included dust dynamics treat the system as being composed of a two-fluid gas+dust, MPI-AMRVAC simulates an entire range of dust sizes simultaneously, allowing us to compare the effect of the KHI on dust of different size and vice versa. Additionally, while many previous gas+dust simulations are focused on a specific astrophysical setup, the results we obtain here are generally valid for many different systems with dust that form shear instabilities. How this is done is outlined in sections \ref{ProbSet} and \ref{FluPro}. In section \ref{LinPha}, we study how the addition of dust changes the linear phase of the KHI as compared to the analytic gas-only solution and the effect of different dust-to-gas density ratios on the growth rate and wavelength dependence of the instability. Further we discuss how different flow velocities change the growth rate of the dusty KHI as compared to the situation with only gas in section \ref{LinPha}. In section \ref{nonLin}, we discuss the dynamics after the linear phase and look at how the distribution of the gas and dust densities are changed due to vortex formation, mixing, and vortex merging. We outline how the KHI strongly alters the distribution of the dust and forms regions devoid of dust and layers where the dust density is increased significantly. We also discuss how the amount of dust in the fluid changes these effects. In section \ref{3DEff}, we discuss in which way 3D effects change the situation in the previous sections. Finally, we compare the results we found for the structure formation of dust in KHI with some findings on filaments observed in molecular clouds and highlight some interesting similarities in section \ref{fila}.

\section{Problem setup}
\label{ProbSet}

\subsection{Numerical method}
\label{NumMeth}

To simulate the KHI, we use the MPI-AMRVAC code \citep{2012JCoPh.231..718K}, a finite volume code suitable for solving any system of hyperbolic partial differential equations. The inclusion of a dust module in the code is discussed in \citet{2011ApJ...734L..26V}. Both gas and dust are treated as separate fluids. To capture species-dependent dust dynamics, multiple dust fluids with different characteristics (e.g.  different grain size or material densities) can be used simultaneously. The gas dynamics are governed by the following equations:
\begin{eqnarray}
	\frac{ \partial \rho}{\partial t}  \quad + & \nabla \cdot (\rho \boldsymbol{v}) & = 0,\\
 	\frac{ \partial (\rho \boldsymbol{v})}{\partial t}  \quad + & \nabla \cdot (\rho \boldsymbol{v} \boldsymbol{v}) + \nabla p & =  \sum_{d=1}^{N} \boldsymbol{f_d}, \label{momentum} \\
	\frac{ \partial e }{\partial t}  \quad + & \nabla \cdot \left[ (p+e) \boldsymbol{v} \right] & =  \sum_{d=1}^{N} \boldsymbol{v}  \cdot\boldsymbol{f_d}, \label{energy} \\
	&  e  = \frac{p}{\gamma - 1} + \frac{\rho v^2}{2},  \label{eq:energy}&  
\end{eqnarray}	
with $\rho$ as the gas density, $\boldsymbol{v}$ the gas velocity, $p$ the thermal pressure, and $\boldsymbol{f_d}$ the drag force per volume as applied on the gas by dust type $d$. The parameter $N$ is the number of dust fluids, $e$ is the total energy density of the gas. Equation (\ref{eq:energy}) describes the two parts contributing to the energy density $e$ - namely, the internal energy with $\gamma$ the adiabatic index and the bulk kinetic energy of the gas. Equation (\ref{momentum}) can be seen as the Euler equation for momentum conservation, with an added term on the right-hand side that represents the total force per volume acting on the gas. Similarly, equation (\ref{energy}) is the Euler equation for energy conservation supplemented with an extra term on the right-hand side which represents the work done by the dust on the gas due to the dragforce. \\
The dust fluids are treated as a pressureless gas: the internal energy of a dust grain only influences the surface temperature of the grain and has no influence on its movement (see \citet{rjl:dust} for a discussion on the dynamics of the pressureless gas equations). Furthermore, dust-dust collisions are typically rare as dust densities are often low in astrophysical fluids (The typical dust particle density in the ISM is $\approx$ 1 m$^{-3}$, or $10^6$ times lower than that of the gas.), and the collisional cross section of grains is small. Even in regions where the dust density is enhanced or comparable to that of the gas density, the particles themselves are still rare compared to the amount of gas particles. Furthermore, dust-dust collisions at typical astrophysical velocities are often inelastic (see e.g. \citet{2009MNRAS.394.1061H} and \citet{2009A&A...502..845O}) and would not contribute to an internal pressure of a dust fluid. For the treatment of the dust fluids, the governing equations are thus
\begin{eqnarray}
	\frac{ \partial \rho_d}{\partial t}  \quad + & \nabla \cdot (\rho_d \boldsymbol{v_D}) & = 0,\\
 	\frac{ \partial (\rho_d \boldsymbol{v_d})}{\partial t}  \quad + & \nabla \cdot (\rho_d \boldsymbol{v_d} \boldsymbol{v_d}) & =  - \boldsymbol{f_d} \label{momentumdust}.
\end{eqnarray}	
Similar to the variables introduced for the gas fluid, $\rho_d$, $\boldsymbol{v_d}$, and $\boldsymbol{f_d}$ are the dust density, bulk dust velocity and drag force of dust fluid $d$, respectively. We have compared our implementation of the gas+dust hydrodynamics with the benchmarks proposed in \citet{2012MNRAS.420.2345L} and find our solutions in agreement with theirs. \\
For the drag force $\boldsymbol{f_d}$, we use a combination of the Epstein drag law for the subsonic regime and the Stokes law for the supersonic regime \citep{1975ApJ...198..583K}, namely
\begin{equation}
	\boldsymbol{f_d} = -(1-\alpha) \pi n_d \rho a_d^2 \Delta  \boldsymbol{v} \sqrt{\Delta  \boldsymbol{v}^2 + v_t^2}, \label{drag}
\end{equation}
with $n_d$ as the dust particle density, $a_d$ the grain radius of fluid $d$, $\Delta  \boldsymbol{v}$ the difference between the gas and dust velocities ($\Delta \boldsymbol{v} = \boldsymbol{v} - \boldsymbol{v_d}$ ), and $v_t = \frac{3}{4} \sqrt{3p/\rho}$ \citep{1975ApJ...198..583K} the thermal speed of the gas. The force is evaluated separately in each spatial direction and the same expression is used independent of the dimensionality of the problem. In this way, the 2D simulations represent a 3D simulation with embedded 3D grains in the limit of vanishing third dimensional velocities and not a truly physical 2D setup with disk-like dust particles, which would represent infinitely long cylinders if expanded to 3D. This way we can directly compare 2D simulations with cut of 3D simulations in the same plane. The variable $\alpha$ is the sticking coefficient and expresses the amount of gas particles that stick to the dust grain after collision. Thus, (1-$\alpha$) is a measure of the effectiveness of momentum transfer from gas to dust and vice versa. Following \citet{1975ApJ...198..583K}, we use $\alpha = 0.25$.\\
Sometimes, it is useful to represent regions as being free of dust, or in some cases, the dust density can become very low due the dynamics. To be able to handle these cases numerically, we introduce a threshold value for minimum density. If the density in a fluid would become lower than the threshold value, the dust from this fluid is removed by setting both the density and the velocity of the fluid to zero. In all simulations described here, the threshold is 10 orders of magnitude below the initial dust density.\\
Several numerical methods for the time advance are implemented in the code, as well as multiple flux limiters. In the simulations described in this paper we used the second order TVDLF (total variation diminishing Lax-Friedrich) scheme in combination with a monotonized central (MC) type limiter \citep{1977JCoPh..23..263V}.\\

\subsection{Numerical setup}
\label{NumSetup}
\begin{figure}
  \centering
  \centerline{\includegraphics[scale=0.5]{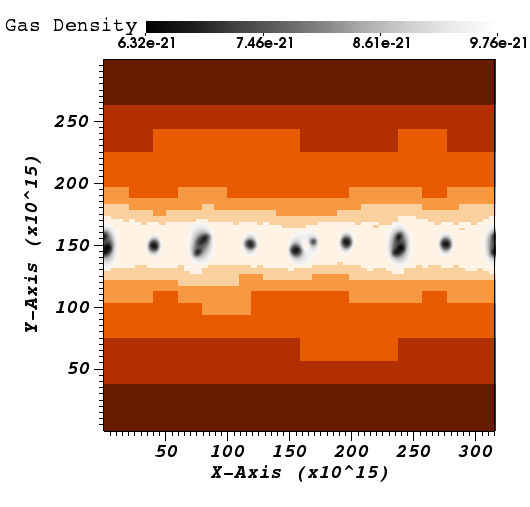}}
  \caption{Full domain of a 2D simulation at t=40. The background colour represents the refinement level with lighter colours being higher levels of refinement. While seven levels are used in the simulation, the coarsed level is no longer used at this time. In the middle, the gas density is overlaid to illustrate how the AMR chooses maximal refinement around the vortices.}
\label{fig:AMRim}
\end{figure}

\begin{figure}
  \centering
  \centerline{\includegraphics[scale=0.55]{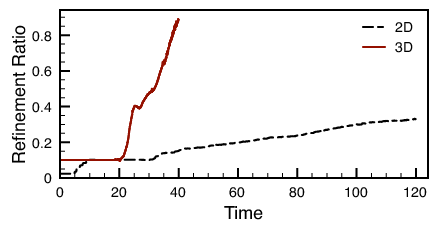}}
  \caption{Evolution of the ratio between the actual amount of cells in a simulation and the amount needed to fill a uniform grid with the same resolution. The difference between the 2D and 3D setup are due to differences in the extent of the simulated domain.}
\label{fig:AMRevol}
\end{figure}

The initial domain we set up for the 2D simulations has a rectangular shape and initially consists of three regions: an upper and lower region with equal velocities, $v_0$, in opposite directions along the $x$-axis, and a thin middle layer that separates them with (total) thickness $D$, where the velocity of the flow varies linearly from the upper to the lower velocity. The existence of the middle layer inhibits instabilities with wavelengths $ \lambda_{min} \lesssim 5D$ for incompressible gas-only KHI \citep{1961hhs..book.....C}. This guarantees that the growth of the KHI is not limited by the smallest resolution scale, which is contrary to a setup of KHI without a middle layer, gravity, and surface tension in which the initial growth increases as $\omega \propto k_x (=2\pi/\lambda)$  for all wavenumbers \citep{1961hhs..book.....C}. The boundaries perpendicular to the $x$-axis are periodic, limiting the development of instabilities of wavelengths longer than the box size; however, as we will typically take a box several times larger than the most unstable wavelength, this will not prove a limitation. Perpendicular to the flow direction, we use open boundaries. Care is taken to assure that these boundaries are always far enough from the instability dominated region during the entire simulation. \\
Depending on the simulation, different initial perturbations have been used. Unless mentioned otherwise, we use a ``smooth'' perturbation of the perpendicular velocity near the middle layer:
\begin{equation}
	v_y = U_0 \exp \left( -\frac{(y-M )^2}{2 \sigma^2} \right) \sin{(k_x x)}
	\label{eq:WavePer}
\end{equation}	
for both the gas and dust fluids. The parameter $U_0$ sets the amplitude of the perturbation and is set as $U_0 = 10^{-3} v_0$ and $M$ is the $y$-coordinate of the middle of the separation layer. To make the perturbation smooth, we take $\sigma = 5D$, which makes the region influenced by the  perturbation several times larger than the middle region. Another approach we used to set up the perturbations is to fill the middle layer with different random $v_y$ values for each cell:
\begin{equation}
	v_y = U_0 \left[ 2\, \textrm{rand}() - 1 \right]
	\label{eq:RandPer}
\end{equation} 
with rand() being a random number between 0 and 1 and $U_0$ having the same value as before. All fluids have the same initial velocities and the same perturbation velocities. \\
In a typical setup, we use seven levels of adaptive mesh refinement (AMR) on a 16$\times$32 base grid, which results in an effective resolution of 1024$\times$2048. The AMR refinements are triggered by a L\"{o}hner type estimator \citep{Lohner}, which computes normalized second derivatives to locate strong variations in specified variables. In all simulations in this paper, the AMR is triggered by a combination of the gas velocity perpendicular to the initial flow (caused initially by the KHI), the gas density $\rho$, and the density of the dust species representing the largest particles (see section \ref{DusFlu}). Figure \ref{fig:AMRim} gives an example of the refinement in a typical 2D simulation. By tracking the time evolution of the actual number of cells as compared to the number of cells needed in a similar simulation without AMR (see figure \ref{fig:AMRevol}), we can compute that the simulations in figure \ref{fig:AMRevol} would take roughly 7.12 and 5.83 times the current time for the 2D and 3D run, respectively. A typical 2D simulation up to $t=120$ uses $\sim$ 200 CPU hours, and the 3D simulation in figure \ref{fig:AMRevol} needs about 15000 CPU hours to complete. More details about the 3D simulation can be found in section \ref{3DEff}.

When setting up the simulation to study the growth of an instability of a certain wavelength $\lambda$, we choose the physical dimensions of the box to be $20\lambda$ in the $x$-direction while keeping the size in the $y$-direction fixed. Doing so, we assure that in every simulation every wavelength of the perturbation is resolved with the same number of cells (viz. $\sim 52$ cells per wavelength), which is independent of the actual wavelength. The aspect ratio of the axes is thus different for each simulation; in terms of code units, the lengths of the $x$ and $y$ axes vary between 0.17$\times$0.3 up to 1.26$\times$0.3. \\

\section{Fluid properties}
\label{FluPro}

The simulations of the dusty KHI are performed with an astrophysical fluid in mind, and the fluid properties have been chosen accordingly. However, the physical quantities in the code itself are scaled to values close to unity, as the Euler equations used to describe the fluid can be written in dimensionless form. Therefore, these simulations are thus also equally relevant for other environments. \\

\begin{table}
\caption{Parameter setup of the physical values used to initialize the simulations. Values are mentioned in code units and in a transformation to physical units.}
\label{tab:par}
\centering
\begin{threeparttable}
\begin{tabular}{l c c}    
\hline\hline 
Parameter              				& Code units 			& Physical units \\
\hline 
Gas Density ($\rho$)					& 10 					& $10^{-20}$ g cm$^{-3}$  \\
Flow velocity$^{a}$ ($v_0$)        	& $5 \times 10^{-3}$ 		& $5 \times 10^{4}$ cm s$^{-1}$  \\
Speed of sound				& $1.19 \times 10^{-2}$	& $1.19 \times 10^{5}$ cm s$^{-1}$ \\
Distance           					& 1 					& $10^{18}$ cm \\
Time	 		     				& 1 					& $10^{11}$ s $\approx 3270$ yr\\
Grain density$^b$	($\rho_p$)	& $3.3 \times 10^{21}$		& 3.3 g cm$^{-3}$\\
Grain size ($a_{min}$-$a_{max}$)	& $(5 - 250)\times 10^{-25}$ 	& 5 nm - 250 nm \\
\hline        
\end{tabular}
 \begin{tablenotes}    
       \item[a] Value of the flow velocity used in most simulations. In section \ref{VelDep} the velocity ranges between $10^{4}$ cm s$^{-1}$ and $10^{5}$ cm s$^{-1}$.
       \item[b] The internal material density inside each dust grain.
     \end{tablenotes}
     \vspace*{0.2cm}
  \end{threeparttable}
\end{table}

\subsection{Gas fluid}
Initially, the gas fluid is uniformly distributed over the entire grid. We use a gas density $\rho = 10^{-20}$ g cm$^{-3}$, or $\sim 6\times 10^{3}$ m$_H$ cm$^{-3}$. The temperature of the gas, which is used for the initial pressure calculation (assuming an ideal gas EOS), is initially set to T = 100 K. In the upper and lower layer, the gas has a velocity $v_0 = 5 \times 10^{4}$ cm s$^{-1}$ in opposite directions (different values are used in section \ref{VelDep}). This is $\sim$ 0.43 times the speed of sound of the medium. We use an adiabatic index of $5/3$, which is appropriate for the used values of gas density and temperature \citep{2013ApJ...763....6T}. A summary of the physical values is given in table \ref{tab:par}.\\

\begin{figure*}
  \centering
  \includegraphics[width=13cm]{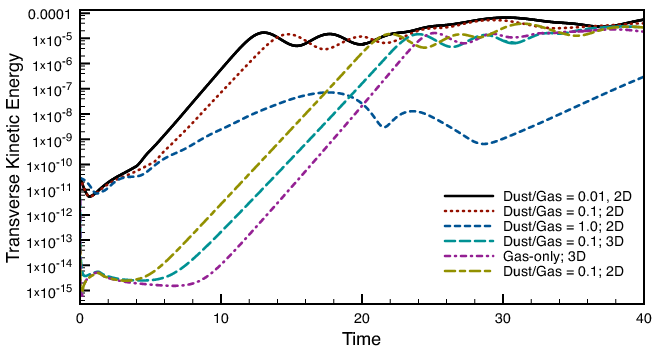}
  \caption{Evolution of the transverse kinetic energy of the gas in six  different simulations with the same dimensionless wavenumber $\kappa=0.7968$ (the most unstable wavenumber in the gas-only case). The vertical axis has been set to logarithmic scale to clearly show the exponential growth in the initial linear growth phase. Three simulations start with a lower initial transverse kinetic energy. This is because they start with small random perturbations in the middle layer, instead of being excited with the perturbation described in equation (\ref{eq:WavePer}). These three simulations are discussed in the section \ref{3DEff}. The behavior of the simulation with $\delta=1$ is different; this is explained in section \ref{WaveDep}.}
\label{fig:Kinetic}
\end{figure*}

\subsection{Dust fluid}
\label{DusFlu}
The total dust density (the sum of the densities of the dust fluids) is chosen by setting the dust-to-gas (mass density) ratio $\delta$. Typically, we use ratios between 0.01 and 1. A canonical value for the dust-to-gas ratio in the ISM of 0.01 \citep{1954ApJ...120....1S} is often used, so the ratios we consider are standard mixtures up to blends which are enriched with dust. The velocities of the dust fluids are initially taken identical to those of the gas fluid, including in the perturbed regions. For each dust species, one has to specify the material density of the dust grains $\rho_p$ (the density of the matter inside a spherical dust grain), which is used to calculate the drag force. Here, we take the dust grains to be composed of silicates, resulting in an internal grain density of $\rho_p$ = 3.3 g cm$^{-3}$ \citep{1984ApJ...285...89D}.  When setting up the dust fluids, a range in dust grain sizes is chosen. We typically choose to include grains of sizes between 5 nm and 250 nm. We then choose the number of bins to represent a certain part of this range in sizes. Each bin is represented as a separate dust fluid during the simulation. In most simulations discussed here, we use four dust fluids; however, this is mentioned explicitly for simulations with another number of subdivisions. Once the number of dust fluids has been chosen, the size range for each bin is chosen by setting an equal density in each bin. This is done by integration over the dust density as computed from the size distribution, the grain volume and grain density:
\begin{eqnarray}
& &\frac{1}{N} \int_{a_{min}}^{a_{max}} n(r) \left[ \frac{4}{3} \pi r^3 \right] \rho_p \,\mathrm{d}r= \int_{R_{min}}^{R_{max}} n(r) \left[ \frac{4}{3} \pi r^3 \right] \rho_p \,\mathrm{d}r \label{BinInt}
\end{eqnarray}	
with ${R_{min}}$ and ${R_{max}}$ as the delimiters of a certain bin, $N$ the number of bins, and $a_{min}$ and $a_{max}$ the minimum and maximum size of the total range, respectively. The function $n(r)$ represents the distribution function for dust particles of radius $r$. We use $n(r) \propto r^{-3.5}$ \citep{1984ApJ...285...89D}, which is a typical size distribution for dust particles in the ISM in equation (\ref{BinInt}) to calculate the upper limit $R_{max}$ of a bin with lower limit $R_{min}$:
\begin{eqnarray}
\Rightarrow \qquad & & R_{max} = \left[ \sqrt{R_{min}} + \frac{1}{N} \left( \sqrt{a_{max}} - \sqrt{a_{min}} \right) \right]^2. \label{BinSize}
\end{eqnarray}	
The ${R_{max}}$ of the first bin can be calculated immediately, further bin widths can be calculated by using equation (\ref{BinSize}) with the width of the first bin. Once the range has been further divided in $N$ bins with for with each bin a (different) width, we need to choose a single grain size $\bar{r}$ for each bin, which is representative for the bin during the simulation. This is done by taking a weighted mean over the bin in question. As a weighting function, one can take the particle size or particle mass; however, we choose a function, which corresponds to the drag force (equation (\ref{drag}), with $a_d = r$):
\begin{eqnarray}
\boldsymbol{f_{d,\Delta r}} = \int\limits_{R_{min}}^{R_{max}} \boldsymbol{f_{d}} \,\mathrm{d}r = -(1-\alpha) \pi \rho \bar{r}^2 \Delta  \boldsymbol{v} \sqrt{\Delta  \boldsymbol{v}^2 + v_t^2} \int\limits_{R_{min}}^{R_{max}} n(r) \,\mathrm{d}r, \label{ForceEq}
\end{eqnarray}
viz. the total force on the bin with range from ${R_{min}}$ to ${R_{max}}$, called $\boldsymbol{f_{d,\Delta r}}$, is the integral over all the grain sizes in the bin. This total force is supposed to be equal to an equivalent force in which the particle size is replaced by the representative size $\bar{r}$. Thus, isolating $\bar{r}$ in equation (\ref{ForceEq}), canceling all parameters outside the integrals, and assuming the same particle size distribution as before gives us
\begin{eqnarray}
	&\bar{r}^2 =& \frac{\int\limits_{R_{min}}^{R_{max}}r^2 n(r) \,\mathrm{d}r}{ \int\limits_{R_{min}}^{R_{max}} n(r) \,\mathrm{d}r} \\
	\Rightarrow \qquad & \bar{r} = & \sqrt{ 5 \frac{R_{max}^{-0.5} - R_{min}^{-0.5} }{ R_{max}^{-2.5} - R_{min}^{-2.5}}}.
\end{eqnarray}
This method assures us that the drag force on all grain sizes is correctly represented by a force on the representative grain size $\bar{r}$ in the bin. An overview of the resulting representative grain sizes for different amounts of dust fluids is given in table \ref{tab:DustSize} in the appendix. \\
All dust fluids interact with the gas fluid through the drag force (equation (\ref{drag})); however, dust-dust interactions are not included. One can demonstrate that the mean free path of a dust grain, $l$, can be written as 
\begin{equation}
	l = \frac{\rho_{p} a_d }{3 \sqrt{2} \rho_d}
\end{equation}
with $\rho_p$ the internal grain density. The mean free path increases with grain size for a certain dust density $\rho_d$. For the values of $\delta$ and $a_d$ mentioned above, the mean free path ranges between $5.9 \times 10^{13}$  cm and $2.9 \times 10^{17}$ cm, while dust-gas collisions only have a mean free path between $2.4 \times 10^5$ cm and $6.0 \times 10^8$ cm. Thus, the exclusion of dust-dust interactions is a valid assumption. As mentioned earlier, the inclusion of dust-dust interactions would involve taking into account physical phenomena, such as dust coagulation and shattering, which are not of the essence here but will be included in the code in the future.\\
In appendix \ref{appA}, a demonstration of the validity of the approach described in the previous section is given by comparing the structures and properties of setups with different amounts of dust fluids.

\section{Results}

\subsection{Linear phase}
\label{LinPha}

\begin{figure}
  \centering
  \centerline{\includegraphics[scale=0.5]{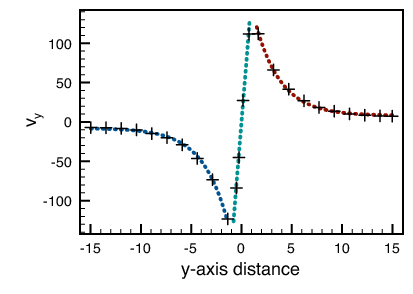}}
  \caption{Comparison of the values of $v_y$ of a simulation ($+$ symbols) and the derived eigenfunctions (dotted lines) on a cut perpendicular to the flow direction. For clarity, only a limited amount of simulated points is shown, The values of $v_y$ are given in cm s$^{-1}$. The cut, taken at $t=3$,  only spans a small portion of the range along the $y$-axis, and is centred on the middle of the boundary layer. The distance from the centre is given in units of $D$. The dotted lines each represent a solution of the eigenfunction, the red line in the upper flow, teal in the middle layer, and blue in the lower flow. From the correspondence with the simulated values, it is clear that during the perpendicular velocity the linear phase evolves from the perturbation in equation (\ref{eq:WavePer}) to the distribution expected from the analytical derivation of the eigenfunction.}
\label{fig:eigenfunc}
\end{figure}

To quantify the growth of the instability through time in the simulations, we use the total kinetic energy of the gas in the $y$-direction (viz. normal to the initial background flow in 2D and 3D) as an indicator. From this measurement, we see how the instability in the boundary layer grows as free energy from the unstable setup is converted in movement perpendicular to the background flow. Several examples of the evolution of the transverse kinetic energy can be seen in figure \ref{fig:Kinetic}, where six evolutions are compared. In the figure, we see how the kinetic energy starts to increase exponentially after a short settling period. This phase, called the linear phase, ends around $t \sim$ 10-25, depending on the simulation. Each growth rate is calculated from a different simulation by fitting the linear growth phase of the transverse kinetic energy with an exponential. We then adjust the values of the growth rate found by the analytic analysis (see section \ref{WaveDep}) so that it can directly be compared with the growth of the kinetic energy by converting the analytic growth rate of velocity to one of kinetic energy (i.e., There is an extra factor of two as the analytic growth rate is valid for $v_y$ and the kinetic energy is dependent on $v_y^2$.). \\
During the linear phase, the spatial evolution of the velocity perpendicular to the flow, $v_y$, can be found as the corresponding eigenfunction. The two-dimensional form of $v_y$ is described by the Taylor-Goldstein equation \citep{1931RSPSA.132..499T,1931RSPSA.132..524G}, 
\begin{equation}
(v_0 - c) \left[ \frac{\partial^2 \psi}{\partial y^2}  - k^2 \psi \right] + \left[\frac{N^2}{v_0 - c} - \frac{\partial^2 v_0}{\partial y^2} \right] = 0,
\label{TG}
\end{equation}
with $v_0$ as the initial flow as described above; $\psi$ a two-dimensional stream function with $\psi = \frac{\partial v_x}{\partial y}$ and $\psi = \frac{\partial v_y}{\partial x}$; $c$ the phase speed; and $N = \sqrt{-\frac{g}{\rho} \frac{\partial \rho}{\partial y}} $ the Brunt-V\"{a}is\"{a}l\"{a} frequency ($g$ is a (gravitational) acceleration in the $y$-direction). It can be seen that equation (\ref{TG}) can easily be solved for the three parts of the domain and coupled at the interface, as both $N$ and the second derivative of $v_0$ vanish everywhere in our setup. Figure \ref{fig:eigenfunc} shows the comparison between the values of $v_y$ in a simulation with $\delta = 0.01$ at $t=3$ and the eigenfunctions in the three parts, as calculated from the gas-only version of the Taylor-Goldstein equation.

\subsubsection{Wavelength dependence of the growth rate}
\label{WaveDep}

To study the growth of the dusty KHI, we compare the growth rates with those of gas-only KHI. As derived in \citet{1961hhs..book.....C} for an incompressible gas, the gas-only setup with two shear layers at $\pm D/2$ and a boundary layer with total thickness $D$ is unstable between $0<\kappa<1.2785$ with $\kappa = k_x D$, a dimensionless number, and $k_x$ the wavenumber of the perturbation. By solving the analytically known dispersion relation, one can find that the setup is most unstable around $\kappa = 0.7968$. As described in section \ref{NumSetup}, each wavelength is investigated by setting up a simulation with length $20\lambda$ in the $x$-direction. For each wavelength, we simulate four different setups; each setup has a different dust-to-gas ratio $\delta$. Namely, $\delta=0$ (no dust, to compare the simulations with the theoretical result), $\delta=0.01$, $\delta=0.1$ and $\delta=1$. For each setup, we investigate 11 different wavelengths. As mentioned in section \ref{LinPha}, we make quantifications of the growth rate of the transverse kinetic energy from the simulations.  \\

\begin{figure}
  \centerline{\includegraphics[width=\hsize]{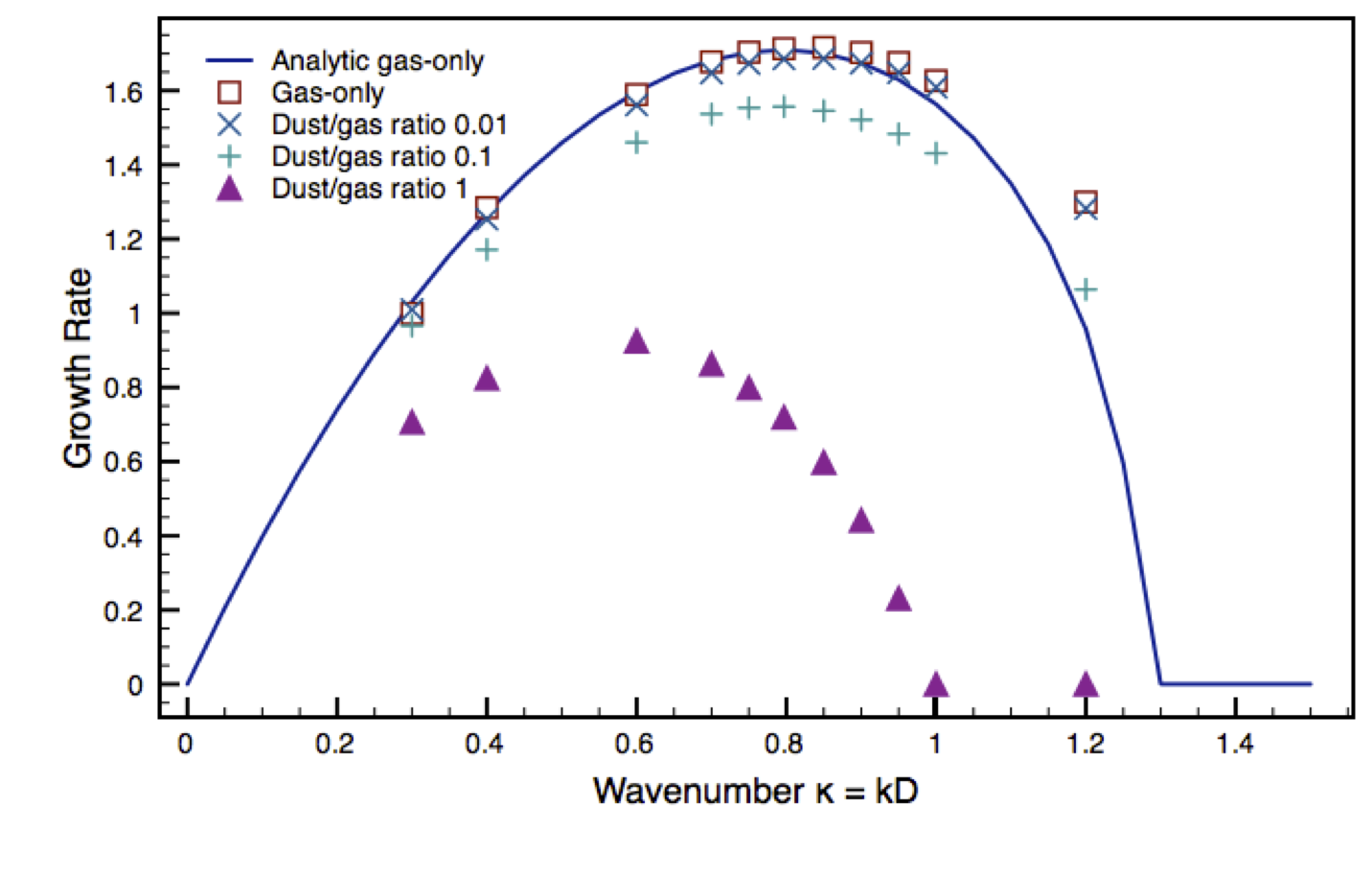}}
    \caption{Dependence of the growth rate of the KHI on the wavelength of the initial perturbation. The continuous line is the analytic solution of the dispersion relation of a gas-only fluid with the setup as described in \ref{NumSetup}. }
\label{fig:Growth}
\end{figure}

The initial transverse velocity perturbation of the middle layer initiates the KHI for the gas fluids, resulting in an exponential growth of the initial perturbation for the gas. Even though the dust fluids are also initially perturbed, they are not prone to the KHI themselves, as the fluids are pressureless. In figure \ref{fig:Growth}, we compare the analytic solution of the dispersion relation with growth rates derived from simulations. It can be seen that the gas-only simulations closely resemble the theoretical curve. The growth rates for wavelengths shorter than the most unstable wavelength (viz. $\kappa$ larger than 0.7968) can be seen to be slightly higher than the theoretical growth for an incompressible gas, which we attribute to finite compressibility effects; the deviation from uniform gas density, $\Delta \rho/\rho$, is typically up to $5-10\%$ near the vortices at the end of the linear phase. Adding dust with a total dust-to-gas ratio of $\delta = 0.01$ only has minor consequences for the growth of the KHI: growth rates are $\sim 2\%$ lower than for the gas-only simulation. We also note that the simulations with dust follow the trend of having raised growth rates compared to the theoretical gas-only curve for higher values of $\kappa$. Increasing $\delta$ to 0.1, we see that the effect of the added dust increases. The reduction in growth rate ranges between $5\%$ and $20\%$ with a stronger decrease for shorter wavelengths. By making a fourth order spline interpolation of the data, we can derive an approximation of the maximum value of the growth rate from the simulations. As noted earlier the growth rate for the gas-only is higher than theoretical for shorter wavelengths, resulting in a derived maximum at $\kappa = 0.81$. For the $\delta = 0.1$ simulations, this maximum shifts slightly to longer wavelengths, having its maximum at $\kappa = 0.79$. Further raising the dust-to-gas ratio to $\delta = 1.0$, we see a strong decrease in growth rates. As seen before for the $\delta = 0.1$ case, the decrease is strongest for the shorter wavelengths. Wavenumbers larger than $\kappa = 1.0$ (or $\lambda > 2\pi D$) are stabilized. The most unstable wavelengths can be found around $\kappa = 0.51$, and growth rates are down by around $50\%$. An example of the slower growth of the $\delta=1$ simulations can be seen in figure \ref{fig:Kinetic}. The simulation shown there is excited with a $\kappa=0.7968$ perturbation, and between $t=6$ and $t=16$, the perturbation grows exponentially. The figure also shows clearly that the growth is significantly slower than in the other simulations. Between $t=18$ and $t=28$, a second settling occurs in which a mode with longer wavelength starts to grow. After $t=28$, a new linear phase starts; this time with a longer wavelength, where we initially set up the simulation to include 20 perturbations with wavenumber  $\kappa=0.7968$. We now see only 12 growing perturbations. This means the wavenumber effectively shifts from  $\kappa=0.7968$ to $\kappa=0.48$ at which the growth rate is faster, as can be seen in figure \ref{fig:Growth}. Note that this is not a physical merger of vortices but the growth of the larger wavelength as it overtakes the growth of the shorter one.\\ The reason for the shift towards smaller wavenumbers can be interpreted by noting that the total inertia of the fluid caused by both dust and gas particles, increases by increasing $\delta$. In contrast, the amount of gas, which causes the KHI, remains the same. Not only is the development of the perturbation velocity in the $y$-direction slowed by the added dust that needs to be dragged along, the added inertia of the dust moving in the background flow means that a longer time is needed to deviate a volume element from a movement in the $x$-direction to a rotational movement in a vortex, and therefore smaller vortices are less effective. This is similar to the effect of a faster background flow, which we discuss in the next section.

\subsubsection{Velocity dependence of the growth rate}
\label{VelDep}

\begin{figure}
  \centerline{\includegraphics[width=\hsize]{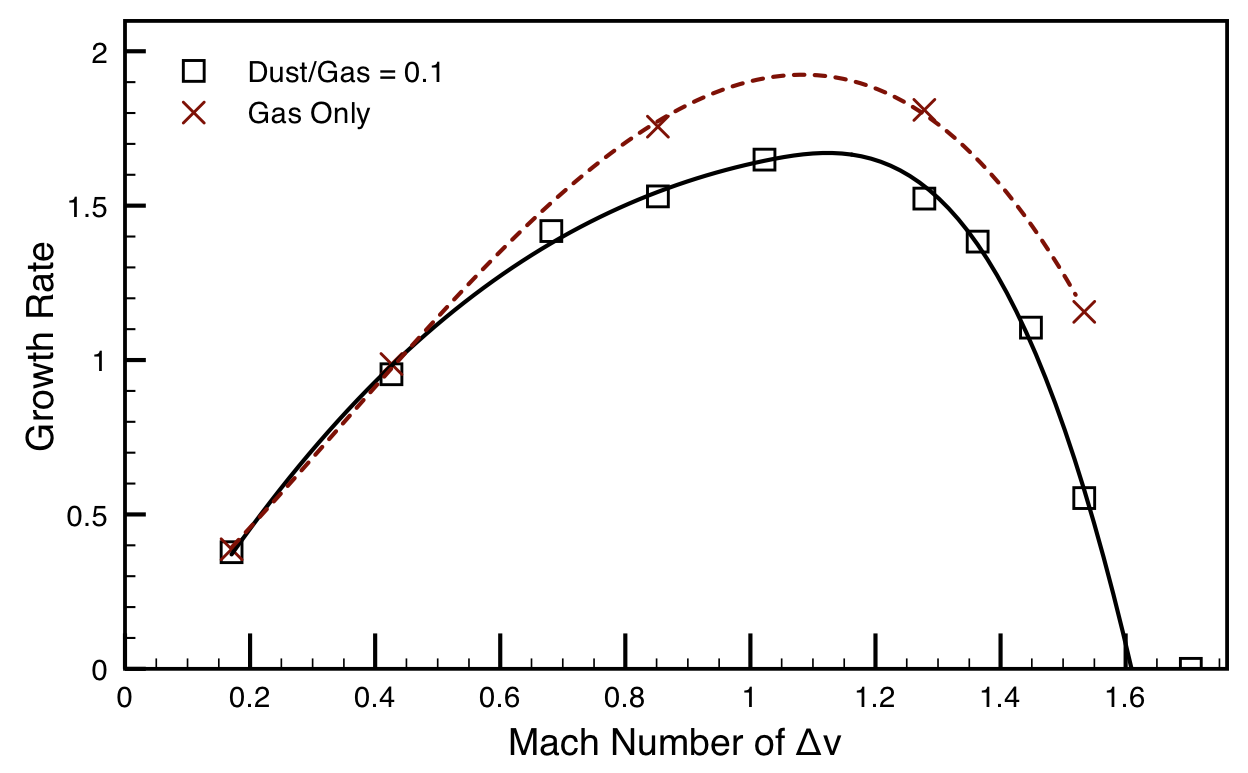}}
  \caption{Comparison of the growth rates of simulations with different fluid velocities. The Mach number indicated is the Mach number of the velocity difference of the upper and lower flow. Fourth order spline fits have been added to guide the eye. Dust tends to stabilize the KHI for all velocities with an increasing effect at higher velocities. The growth rate increases until the velocity difference $\Delta v$ becomes supersonic. }
\label{fig:Mach}
\end{figure}

\begin{figure*}
  \begin{center}
  	\includegraphics[width=5.6cm]{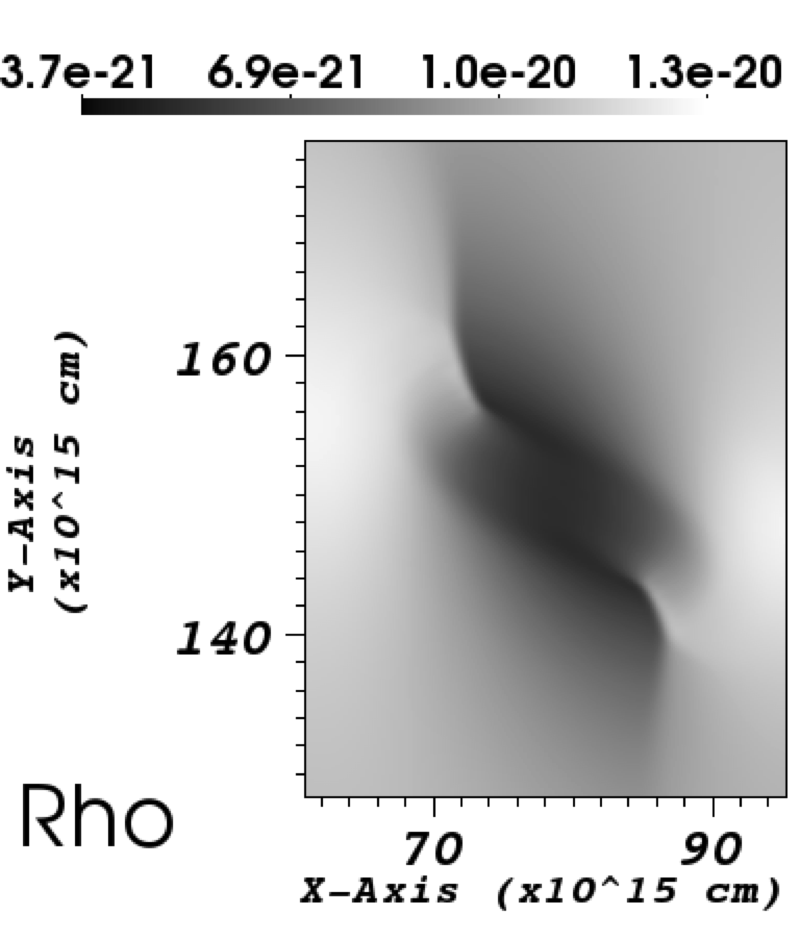} \includegraphics[width=5.6cm]{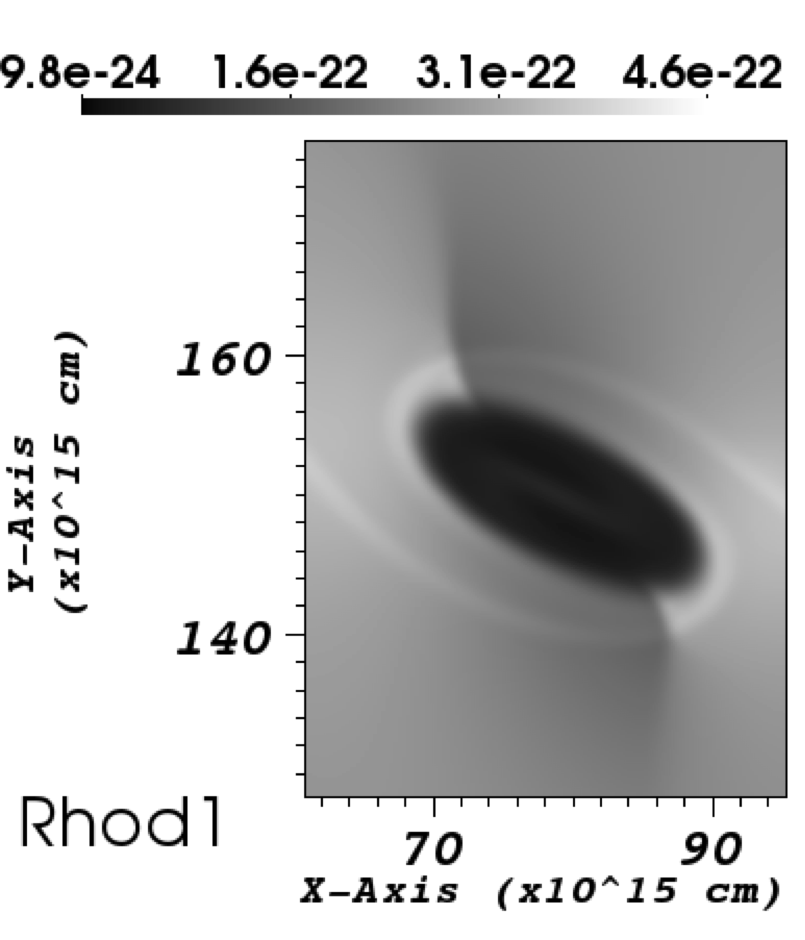}\includegraphics[width=5.6cm]{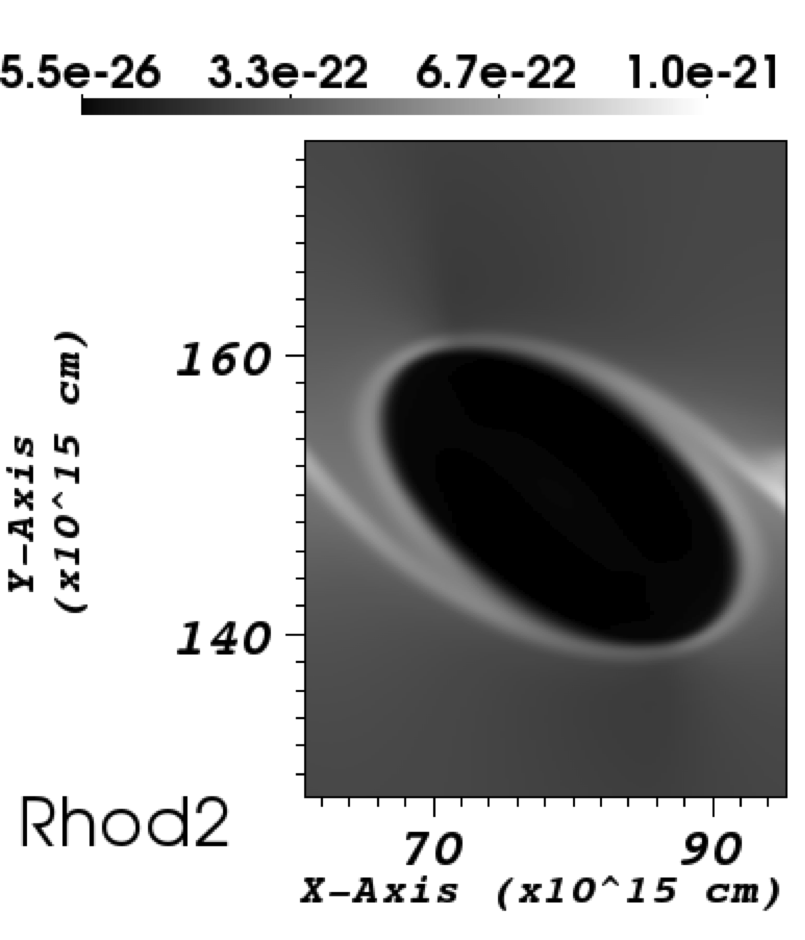}
  \end{center}
  \caption{These vortices have formed shock structures at $t=35$ in a simulation with $\Delta v_0 = 1.51 v_s$, $\kappa=0.7968$ and $\delta = 0.1$. Shocks are clearly visible for the gas (left panel) and the smallest dust species (middle panel). In the second smallest dust species (right panel), the shocks can still be discerned; however, they are less evident. In larger dust species, these shocks are not seen.}
\label{fig:shock}
\end{figure*}

In section \ref{WaveDep} we discussed how the growth rate of the KHI is influenced by the wavelength of the imposed perturbation. Another important quantity that affects the growth rate of the instability is the velocity difference of the two fluids. A comparison between the setup without dust and with dust ($\delta=0.1$) is given in figure \ref{fig:Mach}. In section \ref{WaveDep}, we saw that the growth rate for the $\delta=0.1$ is always lower than for the gas-only case. When we keep the perturbation wavelength fixed (to $\kappa=0.7968$) but change the initial flow velocity $v_0$ we see that the growth rate of the KHI for all values of $v_0$ is once again decreased as compared to the gas-only simulations. For low velocities, growth rates are almost the same with and without dust, but differences increase for larger velocities. The growth rates reach a maximum around $\Delta v_0 = 2v_0 = v_s$ (with $v_s$ the speed of sound). After the maximum, the growth rate decreases, and we observe that the KHI is stabilized when the flow velocity $v_0$ reaches the transonic regime. For supersonic values of $\Delta v_0$, the nonlinear effect ultimately allows the formation of shocks (see figure \ref{fig:shock}), which can be seen to emanate from the vortices (see e.g. \citet{2003PhPl...10.4661B}). We find that these shocks are only seen for fluids with small grains for the dust fluids. This is because small grains are coupled tightly to the dynamics of the gas fluid, while the higher inertia of large grains allows them to cross shocks more easily \citep{2011ApJ...734L..26V}.

\subsubsection{Dust separation}

\begin{figure*}
  \centerline{\includegraphics[width=16cm]{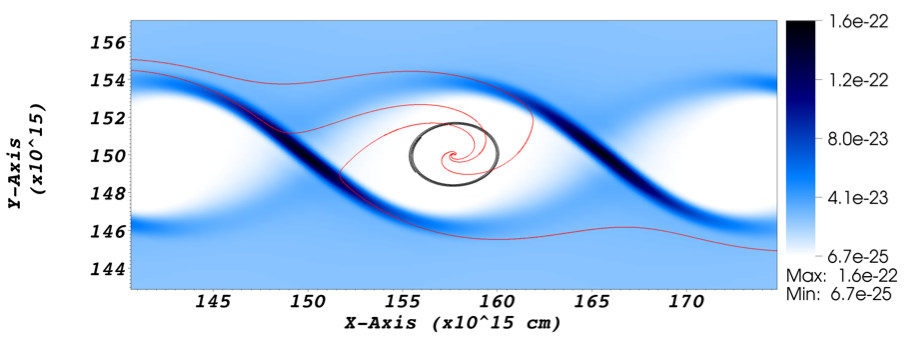}}
  \caption{Density distribution of the largest dust grains out of four species at $t=10$ in a 2D simulation with $\delta=0.01$ and $\kappa=0.7968$. Only part of the simulated domain is shown. In the centre of the middle vortex, the black ellipse is a streamline of the gas fluid. The gas trajectory is (almost) closed as the gas fluid is pressure supported by the lower pressure in the centre of the vortex. By contrast, the streamlines of the large dust particles (shown in red) can be seen to spiral out of the vortex. Ultimately, the dust ends up in two layers below and above the vortex structures. }
\label{fig:Stream}
\end{figure*}

During the linear phase, the typical KH vortices are formed. The gas density distribution is similar to what is typically seen for the gas-only KHI with alternating overdense and underdense islands, corresponding to regions where the pressure has maxima and minima, respectively. In the underdense vortices, the gas rotates forming closed elliptical streamlines (see figure \ref{fig:Stream}). For each vortex, we can study the rotational speed,
\begin{equation}
\omega_{rot} = v_{rot} / r
\end{equation}
by looking at the velocities along several of these streamlines ($v_{rot}$) at multiple distances $r$ from the centre of the vortex. Initially, the rotational speed of the gas in the vortices increases, and the fluids in the vortex have a differential motion: at $t=5$, the rotational speed drops about $50\%$ from around the centre to the edge of the vortex (where the density is the same as the ambient density). Near the end of the linear phase, the rotational speed reaches a uniform value over the entire vortex. After the linear phase the rotation period of the vortices stays constant with a rotation time $T \approx 2\lambda/v_0$. Due to the rotational motion of the gas, the dust is accelerated outwards. As the dust fluids are pressureless, only the interaction with the gas fluid counteracts the outward motion. The smallest dust particles (in our simulation the dust fluid with $\bar{r} \sim 8 \times 10^{-7}$ cm, see table \ref{tab:DustSize}) are more easily stopped (The stopping time of grains due to dust-gas collisions scales linearly with the grain size.) as the particles have lower inertia. Indeed, the stopping time in a simulation with four dust fluids is $\sim$ 24 times larger as that of the smallest ones. This difference in stopping time ensures that the lightest dust is coupled to the gas fluid (the density distribution closely resembles the vorticity of the gas velocity), while heavier dust species are propagated further outward. During the linear phase, an accumulation of dust is observed in the overdense islands.   

\subsection{Non-linear phase}
\label{nonLin}

\subsubsection{Local dust density increase}
\label{LocDustDens}

\begin{figure*}
  \begin{center}
  	\includegraphics[width=8.2cm]{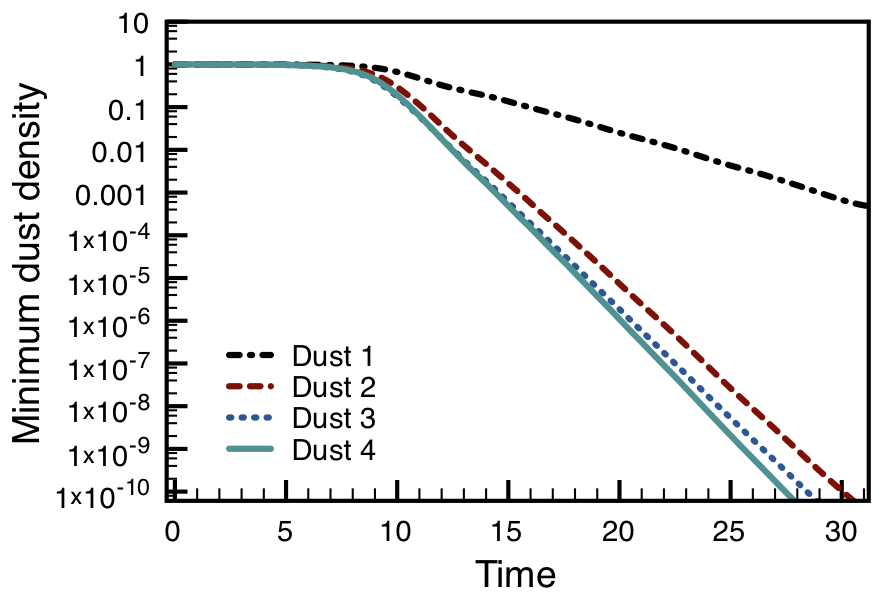} \includegraphics[width=8.2cm]{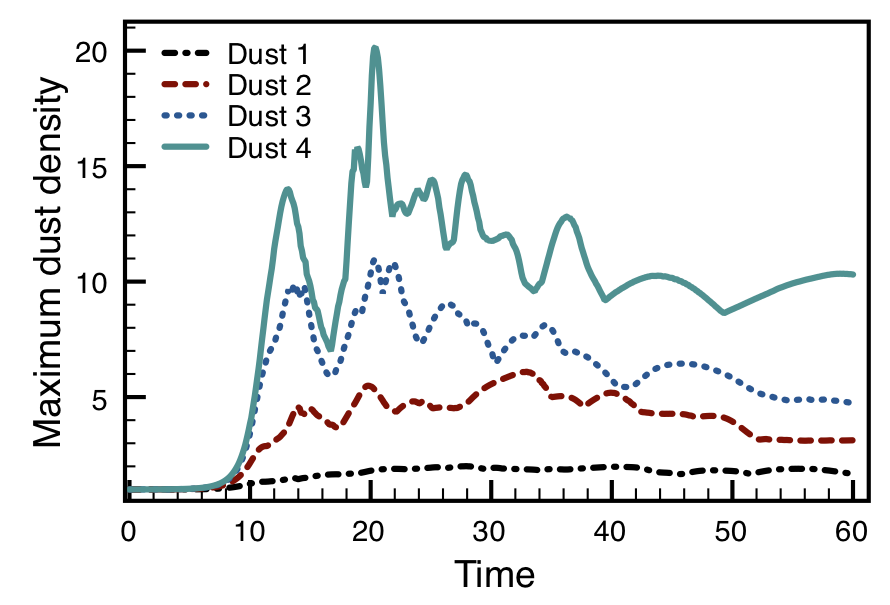}
  \end{center}
  \caption{\textbf{Left}: Evolution of the global minimum in dust density with time for all 4 dust fluids in a simulation with $\delta=0.01$ and $\kappa=0.7968$. The size of the assumed particles in the fluid increases from dust 1 to 4. After the end of the linear phase ($t \sim 10$), the global minimum decreases exponentially. The decrease is faster for heavier dust particles. Only the part of the simulation is shown, where all minima are above the allowed minimum density (see section \ref{NumMeth}). \textbf{Right:} Global maximum for the four dust species. Density increases are stronger for heavier dust species. }
\label{fig:Extrema}
\end{figure*}

\begin{figure*}
  \begin{center}
  	\includegraphics[width=8.2cm]{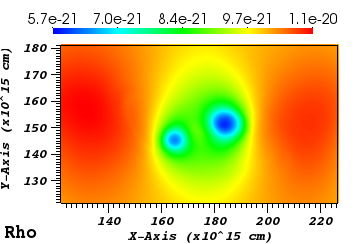} \includegraphics[width=8.2cm]{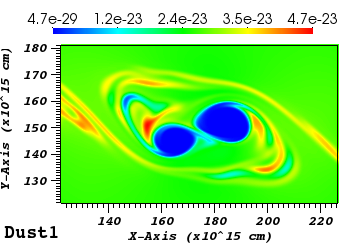} \\ \includegraphics[width=8.2cm]{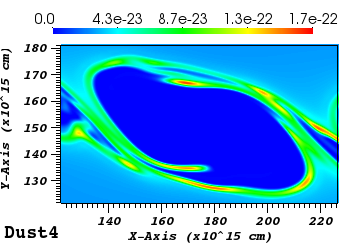}\includegraphics[width=8.2cm]{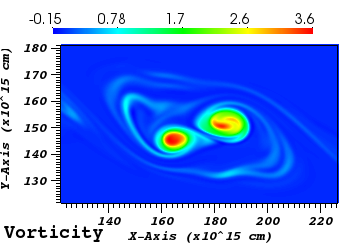}
  \end{center}
  \caption{Density distribution of the gas (top left), the lightest dust species (top right), and the heaviest dust species (bottom left), as well as the vorticity of the gas fluid (bottom right) at $t=60$ in a simulation with $\delta=0.01$and $\kappa=0.7968$. Only part of the total domain is shown. In the image, the merging of two vortices is ongoing. A clear connection can be seen between the vorticity of the gas and the distribution of the smallest dust species. }
\label{fig:comp}
\end{figure*}

In the non-linear phase, which begins around $t \sim 10$ in figure \ref{fig:Kinetic}, the separation of dust from the vortices continues. The heaviest dust species are not stopped during their outward motion and form a layer of increased density around the vortices (see figure \ref{fig:Stream}) by strongly lowering the dust density inside the vortices. If we look at the global minima of the dust fluids (corresponding to centres of vortices) we notice that this quantity decreases exponentially for all the dust sizes after the end of the linear phase (see left figure \ref{fig:Extrema}). The decrease is stronger for heavier particles as they are more easily removed from the vortex due to their higher particle inertia. At the same time, we see that the global maximum of the dust density increases strongly after the linear phase (right in figure \ref{fig:Extrema}). In the beginning, the accumulation is in the regions where the gas is overdense. Later, these regions break up and the maxima of the dust density are located in layers around the vortices. In later stages, vortices become unstable and start to merge. Mergers can also increase local dust density by pushing together dust layers. The dynamical breaking and merging of overdense layers is responsible for the minima and maxima in the right image of figure \ref{fig:Extrema}. Figure \ref{fig:comp} gives a comparison of the gas density with the dust densities of the heaviest and lightest species. The vorticity of the gas is also shown to demonstrate the strong connection between the gas dynamics and the density distribution of the lightest dust species. \\

\begin{figure}
  \centerline{\includegraphics[width=\hsize]{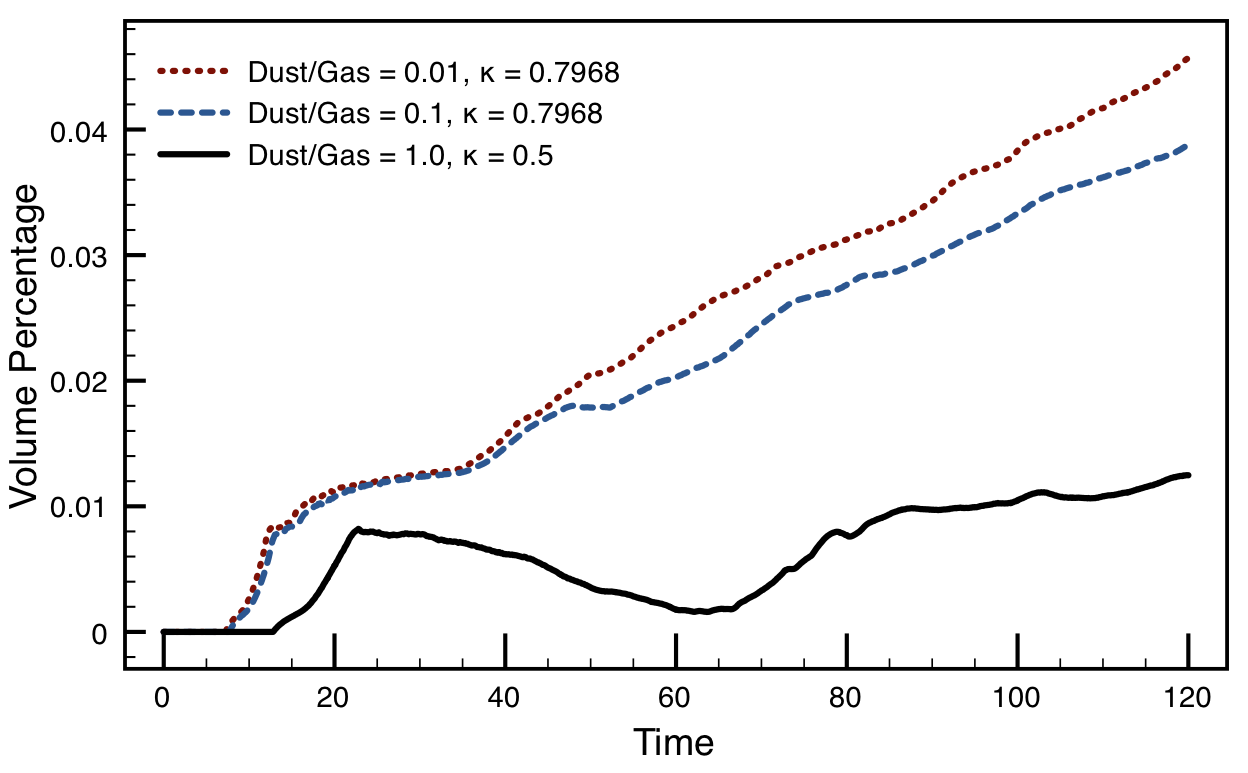}}
  \caption{Volume percentage of the simulated domain where the dust density has at least increased to 2.5 times the initial dust density. All simulations have $v_0 = 0.42$ and are perturbed with wavelengths close to their most unstable value (see figure \ref{fig:Growth}). The portion of the domain where the dust density is significantly enhanced is seen to increase linearly with time after the linear phase ($t \sim 10$ for $\delta = 0.01$ and $\delta = 0.1$ and $t\sim20$ for $\delta \sim 1.0$). The rate of increase decreases with dust-to-gas ratio $\delta$.}
\label{fig:VolPer}
\end{figure}

If we compare setups with different dust-to-gas ratios, we see that setups with lower initial values of $\delta$ have a (slightly) stronger increase in maximum dust density, as compared to the initial density. The maxima and minima of the fluids can, of course, only describe the local behavior in the fluids. For the global density dynamics of the dust fluids we see that this quantity increases linearly during the non-linear phase for both $\delta=0.01$ and $\delta = 0.1$ if we look at the percentage of the volume in which the density has increased by at least a factor of 2.5 (see figure \ref{fig:VolPer}). This trend of linear increase is observed until the end of our simulations at $t=120$ long after the end of the linear phase. The increase is strongest for $\delta = 0.01$ and slightly less strong for $\delta=0.1$. For $\delta=1$, the increase rate is slower. The reason for the observed increase is the merging of vortices. As the vortices merge, the surrounding dust is pushed together, leading to further dust enhancement. This also explains the decreasing growth of volume percentage with dust-to-gas ratio because the dust slows the merging process: the added mass of the dust decelerates the merging motion, which is initiated by the growth of the subharmonic instability of the gas, because it needs to be pulled along by the gas through the drag force, which transfers kinetic energy of the acceleration from the gas to the dust. At the end of the simulation shown in figure \ref{fig:VolPer}, the initial 20 vortices have merged to 2, 3, and 7 vortices for $\delta=$ 0.01, 0.1, and 1.0, respectively.

\subsection{3D effects}
\label{3DEff}

To expand and compare the previous results, we have performed simulations of the dusty KHI in 3D. The setup is comparable to ones in 2D with physical quantities as described in section \ref{FluPro}. The simulation described here has a dust-to-gas ratio $\delta=0.1$. While this value is higher than that typically assumed for the ISM, most conclusions also hold for $\delta=0.01$, as the dynamics in 2D is seen to be largely comparable. Instead of starting with a setup with a fixed wavelength perturbation and a domain with length $20\lambda$, we adopt a simulated domain of $(2\lambda_c)^3$ with $\lambda_c$ as the most unstable wavelength ($\kappa=0.7968$, see section \ref{WaveDep}). In physical units, this translates to a domain of $\sim$ (0.01 pc)$^3$. The smooth perturbation in equation (\ref{eq:WavePer}) is replaced with a random perturbation of $v_y$ in the middle plane. The resolution is set to $256\times1024\times256$, which is four times higher than the resolution perpendicular to the contact plane. One can see that the spatial resolution is actually higher than in the 2D case. To reduce the computational cost, the number of dust species has been reduced to two. The boundaries in the $x$- and $z$-directions are periodic. The boundaries in the $y$-direction (parallel to the contact plane) have open outflow conditions. AMR is used; in the middle region however the grid is forced to the highest refinement during the entire simulation to allow small scale perturbations in the $y$- and $z$-directions to develop from the initial random perturbations. In the surrounding regions, the AMR follows variations in the density of the two dust fluids. While variations in the gas fluid is not traced actively, (because the dust fluid with the smallest dust particles is strongly coupled to the gas fluid) a reasonable grid coverage of the domain of the gas fluid is expected.  \\
We choose to end the simulation at $t=40$, as this value allows us to explore a significant timescale of the non-linear phase. On the other hand, turbulent motions time start filling the simulated domain at this time and increasing refinement (see figure \ref{fig:AMRevol}) causes the simulation time to increase strongly. No stationary state has been reached at this point.

\begin{figure}
  \centerline{\includegraphics[width=\columnwidth]{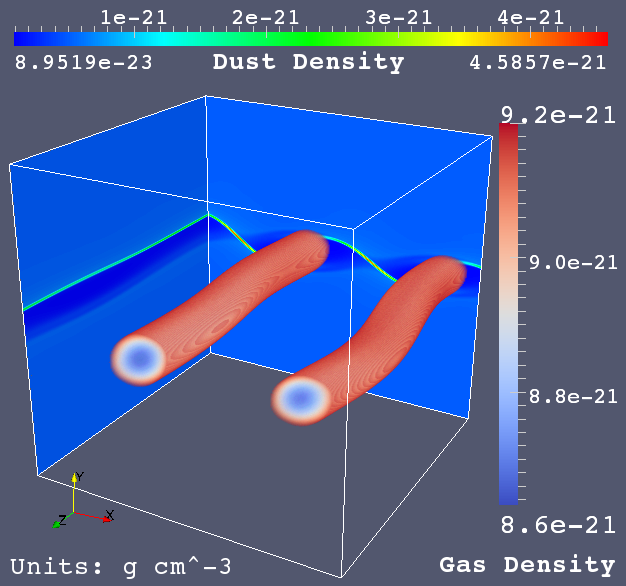} }
  \caption{Volume rendering after the end of the linear phase ($t=22.5$). The entire domain of the simulation is shown. The volume rendering shows the gas distribution of the two straight vortex-tubes that have formed. To visualize the vortices in the volume rendering, we only show gas densities below $9.2\times10^{-21}$ g cm$^{-3}$. At the back and left side of the cube ($xy$- and $yz$-planes, respectively), the dust distribution of the heaviest dust species is shown. The dust forms tubes around the gas vortices, and sheets of increased density between the tubes. }
\label{fig:Tubes1}
\end{figure}

\begin{figure}
  \centerline{\includegraphics[width=\columnwidth]{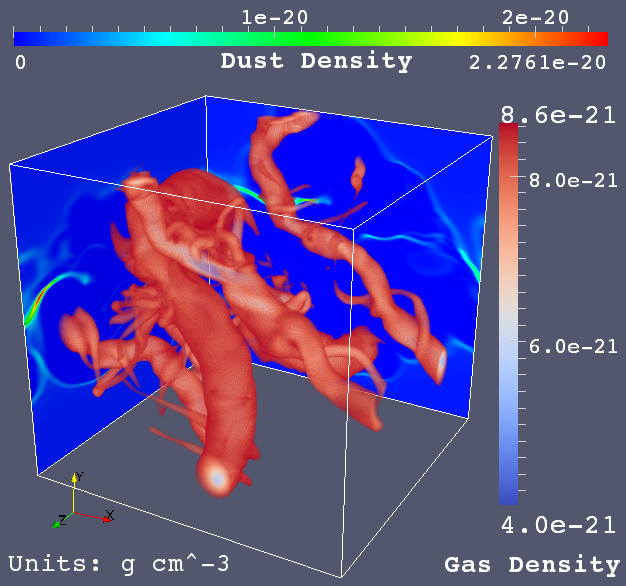} }
  \caption{Volume rendering at $t=40$. To visualize the vortices in the volume rendering, we only show gas densities below $8.6\times10^{-21}$ g cm$^{-3}$. The straight vortex tubes seen at the end of the linear phase in figure \ref{fig:Tubes1} have become severely twisted and are breaking apart. The dust distribution at the same time is shown in figure \ref{fig:3D}.}
\label{fig:Tubes2}
\end{figure}

\begin{figure}
  \centerline{\includegraphics[width=\hsize]{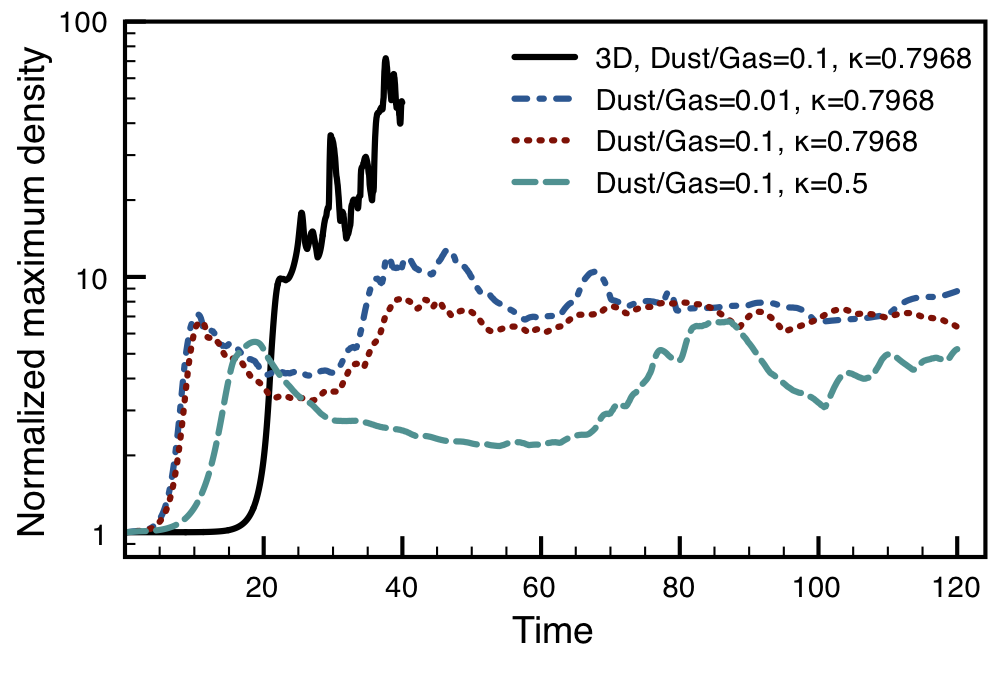}}
  \caption{Comparison of the dust enhancement in four different simulations with different values of $\delta$. The graph shows the maximum of the dust density in the simulation, ascompared to the initial value of the dust density. Dust enhancements are (initially) slightly higher for lower values of $\delta$. The enhancements in the 3D case are significantly stronger due to the additional instability of the vortex tubes in the added third dimension. A log-scale is used for the maximum density. Note that the maximum density starts to increase at a later time in the 3D simulation as compared to the 2D cases because a random initial perturbation is used in 3D.}
\label{fig:DensEnh}
\end{figure}

\begin{figure}
  \centerline{\includegraphics[width=\hsize]{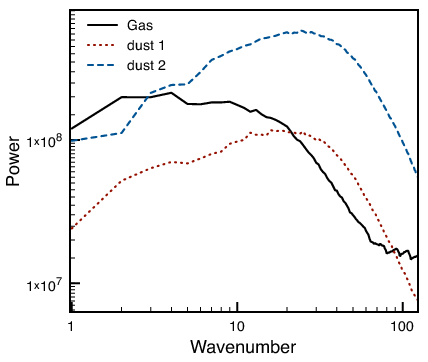}}
  \caption{Power per lengthscale derived from the 3D Fourier transform of the densities of the three fluids at $t=40$. Formed structures are smaller for the dust species, with dust 1 being the dust species representing the smaller dust particles, and dust 2 representing the larger particles. The ripples at high wavenumber for the gas are due to the AMR, as we choose the active refinement to follow the density variations of the two dust fluids.}
\label{fig:FFT3D}
\end{figure}

\begin{figure}
  \centerline{\includegraphics[width=\hsize]{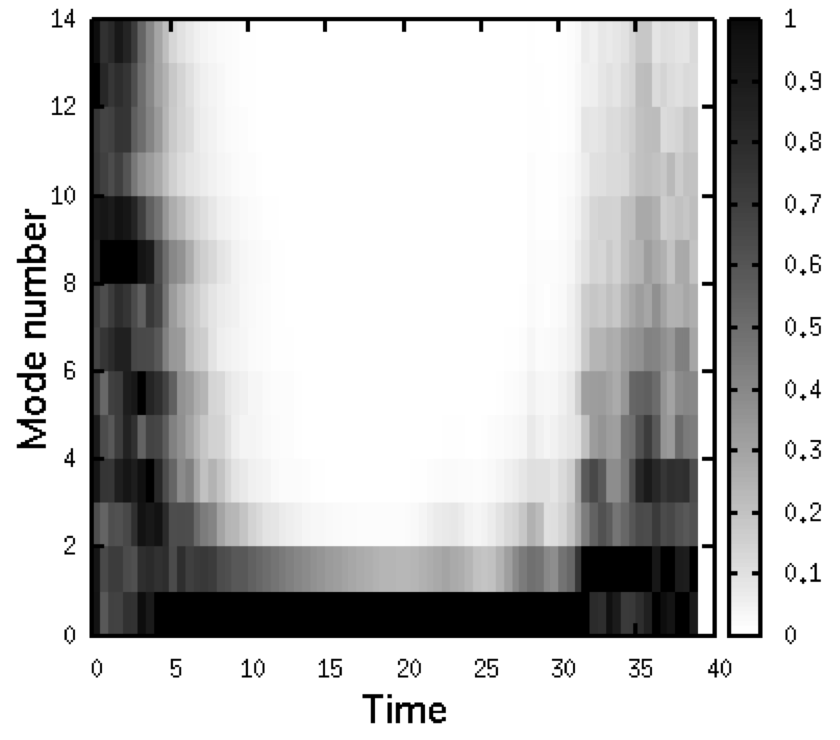}}
  \caption{Comparison of the strength of each mode of $v_y$ along the $z$-direction of the middle plane. Modes 0 to 13 are shown. At each time, the strength of each mode is compared to the strength of the dominant mode at that time. Three distinct regimes can be discerned in the figure. Initially, the initial random perturbations evolve up to $t\sim8$ from small-scale structures to a large scale feature. In the linear phase and the early non-linear phase (between $t\sim8$ and $t\sim30$), the vortex tubes are elongated along the $z$-direction and the 0-mode is dominant. After $t\sim30$, the vortex tubes become unstable, resulting in the formation of small-scale structures.}
\label{fig:Modes}
\end{figure}

To be able to analyze the results in 3D, we have made a 2D simulation with the same setup as described above and for the same simulation in 3D without dust. In the linear part of the instability, the 3D KHI behaves similar to what we have seen in the 2D cases in the previous sections. The observed growth rate in 3D is almost exactly the same as in the 2D case, as can be seen in figure \ref{fig:Kinetic} (i.e. it is $1\%$ slower). Also in the same line as found earlier in 2D, the growth in 3D without dust is $8\%$ faster than with the added dust. As is expected from the setup, initially two vortices are formed. In the linear phase and the early non-linear phase, these vortices are straight, tube-like structures in the added $z$-direction, which adds no 3D-effects to the simulation (see figure \ref{fig:Tubes1}). Similar to the heavy dust layers in 2D, sheets of increased dust density are formed around the vortices and between the vortices in the braids. At around $t=30$, a secondary instability occurs (see figure \ref{fig:Modes}). A discussion of secondary KH instabilities which are inherently 3D effects in the KHI can be found in \citet{1991JFM...227...71K,2000JFM...413....1C} for stratified and homogeneous flows. Here, the three-dimensional instability suppresses the merging mode of the vortices and results in a twisting motion of the vortices, lifting the tubes out of the middle plane and causing a turbulent-like state as discussed in \citet{1987JFM...184..207M}. We also find the secondary instability in the gas-only 3D simulation, as is expected because it is not triggered by the dust itself. The twisting motion causes the dust sheets to be folded and pushed together, leading to the formation of elongated, filamentary dust structures with enhanced dust densities. If we compare the dust density enhancement of simulations with different values of $\delta$ with the enhancement in the 3D simulation (figure \ref{fig:DensEnh}), a clear difference can be seen. Where in the 2D cases, enhancements up to a factor of 10 are typical; the folding of the dust sheets due to the bending of the vortex tubes creates a significantly stronger increase. Dust clumping in 3D is up to almost an order of magnitude more efficient as compared to the 2D cases. In the 2D simulation with the same configuration as that of the 3D one, the enhancement factors are similar to those in other 2D simulations, reaching a maximum enhancement of ten times the initial density dust. This confirms that the increased maximum increase is inherently due to 3D effects.\\
Eventually, the twisting of the vortex tubes leads to a disruption of the vortex structures, as can be seen in figure \ref{fig:Tubes2}. This is in contrast to the 2D case, where an evolution from small to large scales is typically seen. The formation of small-scale dust structures with enhanced densities can also clearly be seen in the power spectrum of the simulation. If we take a 3D FFT (fast Fourier transform) of the density distributions of the gas and dust fluids at $t=40$ (the end of the simulation) and convert this to give the power per lengthscale as is shown in figure \ref{fig:FFT3D}, we clearly see how the maximum in the spectrum shifts to smaller scales if we go from the gas fluid (wavenumber $\kappa \sim 4$, tracing the size of the vortices) to the smallest dust species ($\kappa\sim17$) and to even smaller scales if we go to the largest dust species ($\kappa\sim24$). Figure \ref{fig:Modes} gives a quantitative comparison of the strength of the 13 first modes in the middle plane as a function of time. It clearly shows the evolution to straight vortex tubes during the linear phase, followed by an early non-linear phase where there are no important 3D effects and ultimately the disruption to a turbulent phase in which the vortices break up into small-scale structures.


\section{Discussion: Filaments in molecular clouds}
\label{fila}

\begin{figure*}
  \begin{center}
  	\includegraphics[width=.85\columnwidth]{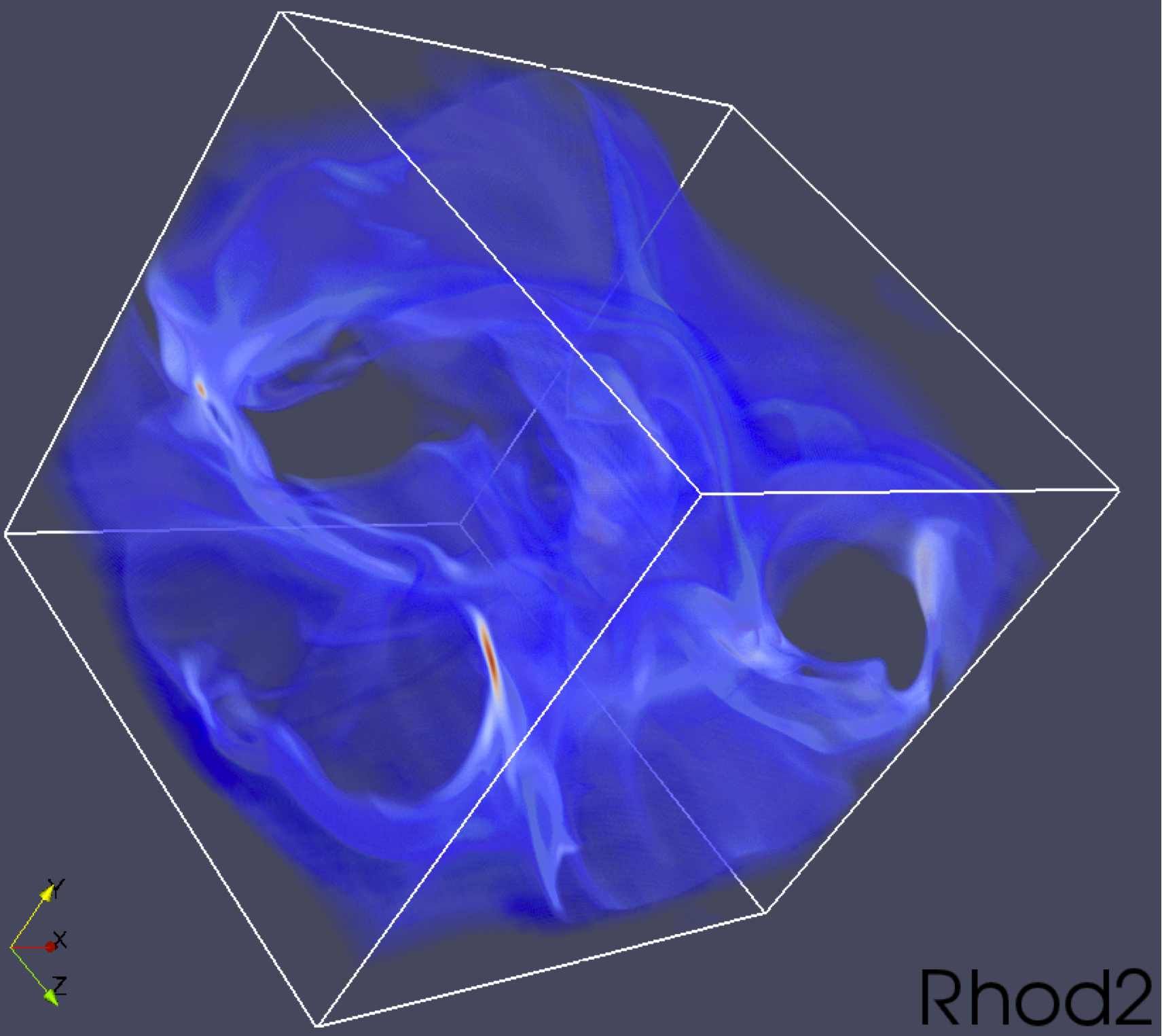} \includegraphics[width=.85\columnwidth]{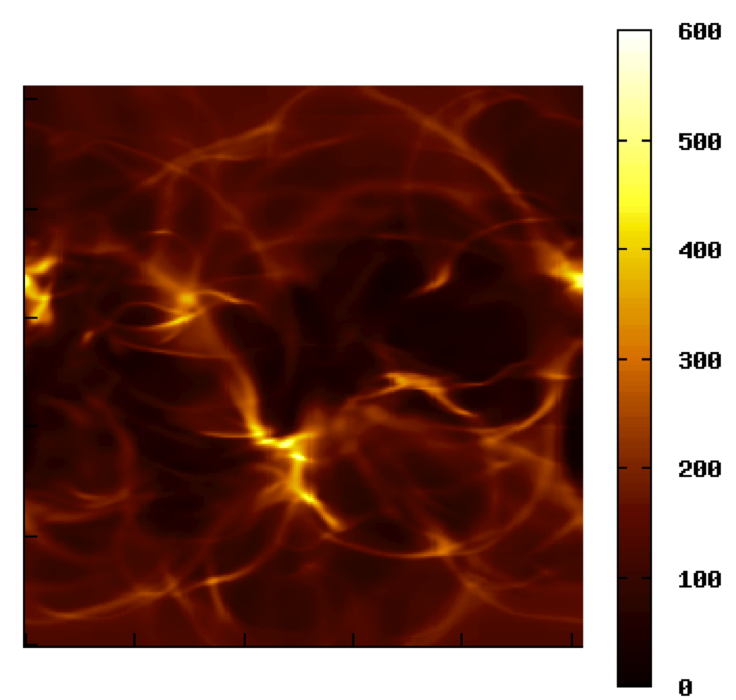}
  \end{center}
  \caption{\textbf{Left:} Volume plot of a simulation of the KHI at t = 40. The heaviest dust species (of two species) is shown with the opacity in the plot scaled with the dust density. One can clearly see the elongated, filamentary structures with several high-density clumps where the density has increased to more than 10 times its value. In between, the structures are vacuum bubbles where the density has decreased by several orders of magnitude. \textbf{Right:} Line of sight surface density of the heaviest dust species, as derived from the 3D simulation on the left. The density scale is in code units. Elongated structures with embedded high-density clumps are clearly seen.}
\label{fig:3D}
\end{figure*}

\begin{figure}
  \centerline{\includegraphics[scale=0.5]{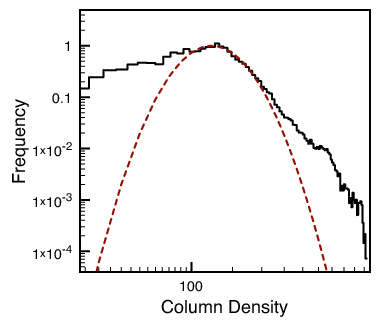}}
  \caption{Column density distribution of the synthetic observation of the KHI on the right side of figure \ref{fig:3D} at $t=40$, as shown in black. The column density is shown in code units. The central part of the distribution is fitted with a log-normal distribution, as shown as the dashed red line. A power-law tail is seen on the high-density side, as well as an additional excess on the low-density side, which is comparable to what is seen in some observations of molecular clouds.}
\label{fig:PDF}
\end{figure}

The simulations described above all have physical parameters, which are similar to values typically observed in molecular clouds. In this section, we highlight some similarities between the simulated KHI in 3D and structures observed in molecular clouds. Molecular clouds are regions in interstellar space where the local density is much higher than in the surrounding regions. Also temperatures are typically much lower.  The mean diameter of a giant molecular cloud is 45 pc \citep{1993prpl.conf..125B}. Structure formation is observed on all sizes in molecular clouds from 0.003 to 30 pc \citep{1991mocl.conf...75G}. Typical substructures are called ``clumps'', ``sheets'', ``bubbles'' and ``filaments''. We have seen in section \ref{3DEff} that these types of structures are also formed in the 3D KHI (as can be seen in the left panel of figure \ref{fig:3D}). Due to the high densities and low temperatures, observations of molecular clouds are often made in infrared wavelengths, in which the emission traces the dust distribution. To compare our 3D simulation with observation, we make a synthetic observation by summing the total dust density along the line of sight in a certain direction, giving us the dust mass column density $\sigma_{d}$:
\begin{equation}
\sigma_{d} = \int \rho_d \,\mathrm{d}s = \frac{1}{\kappa_{\nu}} \int \kappa_{\nu} \rho_d  \,\mathrm{d}s = \frac{\tau_{\nu}}{\kappa_{\nu}}, 
\end{equation}
with $\kappa_{\nu}$ as the dust opacity and $\tau_{\nu}$ the optical depth at frequency $\nu$. If we assume that we look at the system in a wavelength at which it is optically thin ($\tau_{\nu} \ll 1$), the radiation is caused by thermal dust emission and the dust temperature is uniform, then $\sigma_d$ can indeed be used as an estimator for the observed flux of the system. The radiative transfer then simplifies to
\begin{equation}
I_{\nu} \approx B_{\nu}(T_{dust}) \tau_{\nu} = B_{\nu}(T_{dust}) \kappa_{\nu} \sigma_d,
\end{equation}
and thus the emitted intensity scales linearly with the mass column density. The $\sigma_d$ is calculated in the right panel of figure \ref{fig:3D} from the 3D data of the left panel. This total density along the line of sight can then be compared to column densities, which are derived from the observations. One of the important measurements derived from the column density is the probability density function (PDF), which is the probability distribution of the column density values in the observation. It has been shown that the PDF of isothermal structures with turbulence and negligible magnetic and gravitational forces is taking the shape of a log-normal distribution \citep{1994ApJ...423..681V}. This result has also been confirmed from magnetohydrodynamics-simulations with supersonic turbulence and self-gravity, \citep{2001ApJ...546..980O} and isothermal hydrodynamical simulations with supersonic turbulence without self-gravity \citep{2010A&A...512A..81F}. This log-normal shape of the PDF is also seen in observation; however, often a power-law excess is seen on the side of the higher column densities \citep{2009A&A...508L..35K}. \citet{2009A&A...508L..35K} find evidence that links this possible excess to molecular clouds with active star-formation.
Significant deviations from log-normal behavior is also seen in the low-density regions of many clouds.
While the log-normal distribution is generally seen as a consequence of supersonic turbulence, \citet{2010MNRAS.408.1089T} show that such distributions are obtained without supersonic turbulence in three very different numerical setups of molecular clouds. Furthermore, they claim that the development of power-law tails is not due to intermittency or gravity taking over the dynamics.\\
If we make a PDF from our synthetic dust observation as shown in the right side of figure \ref{fig:3D}, we get the distribution shown in figure \ref{fig:PDF}. We thus see that the PDF in our simulation of the KHI has evolved from one initial  value of the column density to a distribution spanning two orders of magnitude. The central part of the PDF is fitted by a log-normal distribution; a power-law-like excess is seen on the high density side, as well as an excess on the low density side. This is very similar to some of the observations in  \citet{2009A&A...508L..35K}. The excesses can be explained in the case of the KHI; values on the low density side correspond to the low-density regions created by the vortex formation in the KHI. The higher density regions are the structures of enhanced dust density formed first by the dust separation in the non-linear phase of the KHI, and later in the non-linear phase by the instability in the third dimension. \\
While the other mechanisms that we discussed above, which have been studied previously, may indeed be important for the formation of the structures observed in molecular clouds, it seems reasonable that the omniprecense of shear flows in astrophysical fluids can also contribute to the formation of the structures and to the formation of the log-normal distribution in the PDFs. This may explain some of the deviations from the log-normal distribution. In future work, we explore this further by performing dedicated simulations with properties derived from observations and link these with the dust radiative transfer code SKIRT \citep{2011ApJS..196...22B} to be able to conduct a more detailed comparison with observations.

\section{Conclusions}

We have presented the wavelength dependence of the growth rate of the KHI with dust-to-gas ratios which range between $\delta=0$ and $\delta=1$. To do so, we performed 44 individual simulations with different values of $\delta$ and $\kappa$. We see that the growth rate reduced slightly ($2\%$) for $\delta=0.01$. Adding more dust reduces the growth rate more significantly. The reduction is stronger for short wavelengths; for $\delta=1$, short wavelengths are completely stabilized. This leads to a shift of the most unstable wavelength to longer wavelengths. For the linear phase of the KHI, we have also investigated the dependence of the growth rate on the shearing velocity from the case $\delta=0.1$. We see that the difference with gas-only simulations is stronger for increasing shearing velocities. Shocks are seen in the gas when the velocity difference becomes supersonic; for the dust, these shocks are only seen for small grains, as the larger grains are more weakly coupled to the gas.\\
The formation of the vortices evacuates dust from the low-pressure vortices at a fast rate, with the minimum density decreasing exponentially for all dust species after the linear phase. High density dust layers are formed around the vortices and in the braids between vortices. Small dust particles form structures closer to the centre of the vortex. The formation of both low density and high density dust regions is stronger for increasing dust particle sizes in the range between 5-250 nm investigated here. The volume percentage of the regions, which have strongly increased dust densities, increases linearly after the end of the linear phase for the cases $\delta=0.01$ and $\delta=0.1$. For $\delta=1$, the trend is less clear. In 2D simulations, where we start with 20 vortices, merging leads to an increase in the size of the structures. In our 3D simulations, the vortex tubes become unstable in the additional third direction, which leads to a breakup of the tubes and eventually to the formation of small-scale structures. These structures are smaller for larger particles than for smaller particles or gas. The additional instability also increases the enhancement of dust densities even further with peaks in density enhancement up to seven times stronger than in comparable 2D simulations.\\
We see that a dusty KHI with physical values comparable in molecular clouds is able to form structures such as filaments, which are  prevalent in molecular clouds, in relative short amounts of time (the end of the 3D simulation at $t=40$ corresponds to 0.13 Myr). Furthermore, the KHI reshapes the PDF of the column densities from a single value in the initial uniform distribution to a central log-normal distribution with excesses on both sides. This is comparable to what is seen in observations without needing supersonic turbulence or self-gravity to form the log-normal distribution and the power-law tail. \\

\begin{acknowledgements}
We acknowledge financial support from the EC FP7/2007-2013 grant
agreement SWIFF (no. 263340) and from project GOA/2009/009 (KU Leuven).
This research has been funded by the Interuniversity Attraction Poles
Programme initiated by the Belgian Science Policy Office (IAP P7/08
CHARM). Part of the simulations used the infrastructure of the VSC -
Flemish Supercomputer Center, funded by the Hercules Foundation and the
Flemish Government - Department EWI. The authors would like to thank the anonymous referee for the valuable comments.
\end{acknowledgements}

\bibliography{KHPaper}

\appendix
\section{Method Validation}
\label{appA}

\begin{figure*}
  \begin{center}
  	\includegraphics[width=5.6 cm]{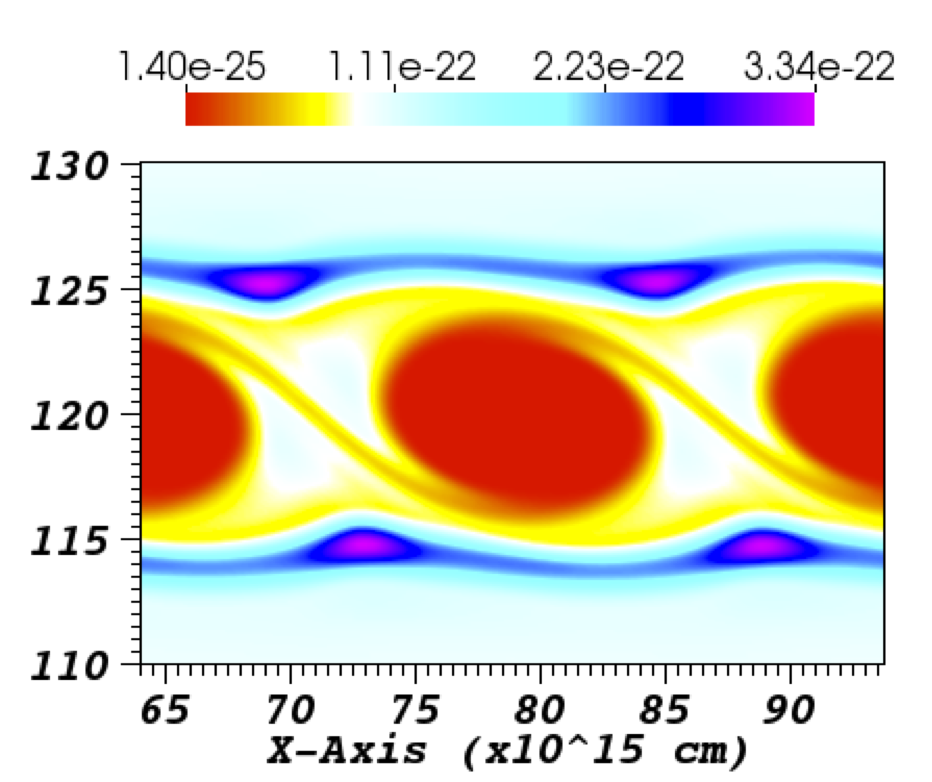} \includegraphics[width=5.6 cm]{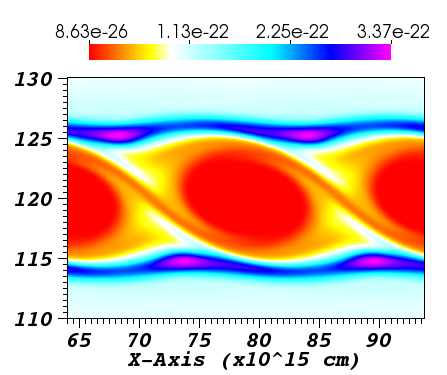}\includegraphics[width=5.6 cm]{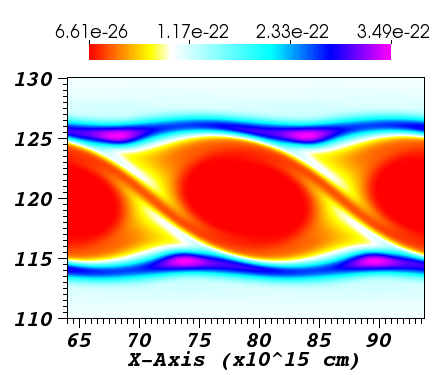}
  \end{center}
  \caption{The summation of the densities for all the different dust species gives us the total dust density, as shown here for a simulation with two dust species (left panel), four dust species (central panel), and eight dust species (right panel). Only part of the larger domain of (1.58$\times$2.4) $\times 10^{17}$cm is shown. While the structures in the simulations with four and eight species are similar, the simulations with only two dust species display a less diffusive structure. The rarefaction rate is also underestimated in the simulation with two species, leading to higher minimum densities than in the two other cases. This is due to lower values of the largest represented dust species. }
\label{fig:diff}
\end{figure*}

\begin{figure}
  \centerline{\includegraphics[scale=0.45]{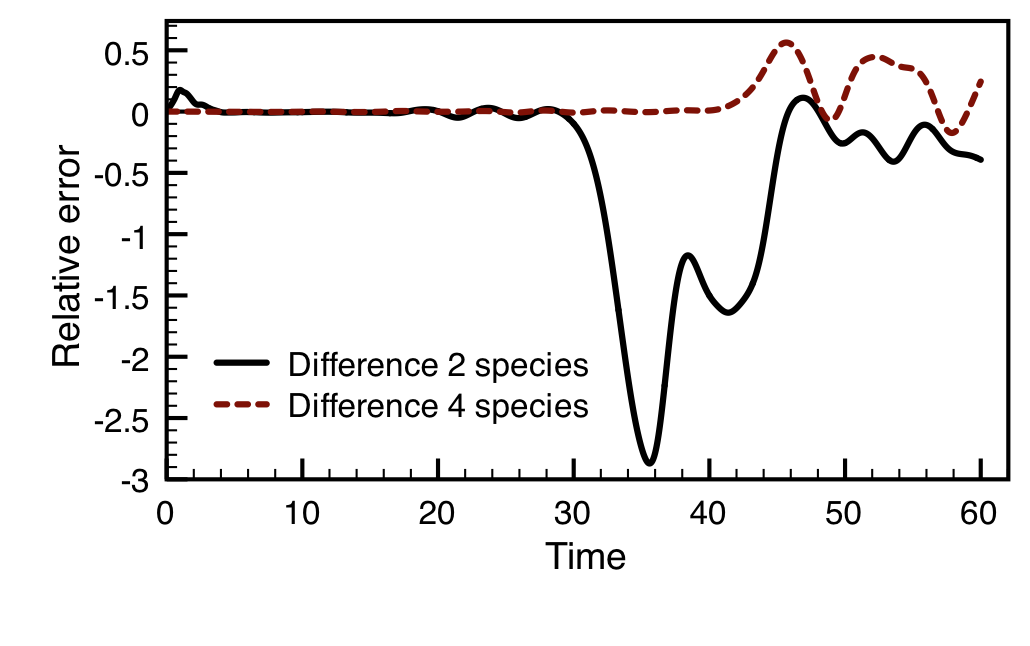}}
    \caption{Relative error of the transverse kinetic energy in simulations with two and four dust species as compared to the simulation with eight species. }
\label{fig:error}
\end{figure}

To test the effectiveness of the dust treatment described in section \ref{DusFlu}, we compare between three simulations with the same physical parameters (see table \ref{tab:par} with $\delta=0.01$ and a total domain of (1.58$\times$2.4) $\times 10^{17}$cm), however with different amounts of dust fluids; namely two, four and eight different species with sizes between 5nm and 250nm. The effective sizes $\bar{r}$ for each fluid, calculated using equations (\ref{BinInt}) and (\ref{BinSize}), are given in table \ref{tab:DustSize}. For all three setups, simulations are conducted up to $t=60$, where non-linear effects, such as vortex merging, are observed. \\
Most importantly, no large differences are seen between the behavior of these three simulations. Their linear phase is equally long, and in the non-linear phase, the vortices merge at the same rate and form structures of comparable sizes. However, small differences can be seen: if we look at the dust distribution in the vortices (see figure \ref{fig:diff}), we see that the total dust density is more spread out for the simulation with four and eight species compared to the other setup. In simulations with more fluids, the outer bins have smaller/larger representative particle radii, leading to more extreme behavior. As different dust sizes are evacuated from the vortex at different rates, this allows simulations with more fluids to have more distinct behaviors, leading to more diffuse distributions of the total dust density. The minimum values are also slightly lower for simulations with more fluids, an effect of having larger values for the maximum effective grain radius, as discussed in section \ref{LocDustDens}.
More quantitatively, we compare the evolution of the transverse kinetic energy of the three simulations. An example of the typical evolution of the transverse kinetic energy is shown in figure \ref{fig:Kinetic}. Here, we look at the relative error of the simulations, as compared to the setup with eight species (figure \ref{fig:error}). For the simulation with four species, the difference is less than 2\% until $t\approx 40$, which is long after the end of the linear phase ($t=10$). Even in the non-linear phase differences are never bigger than 50\%. The differences for the simulation with two species are larger, with a difference of $\sim20\%$ at the beginning of the simulation and a relative difference up to three times the value of the simulation with eight species in the non-linear phase. Nonetheless, differences between slightly different setups are expected due to the non-linear nature later in the simulations. \\

\begin{table}
\caption{Values of the representative dust size $\bar{r}$ of each species, as used in simulations with two, four and eight dust species. The values represent dust sizes between 5nm and 250nm, and have been calculated with equations (\ref{BinInt}) and (\ref{BinSize}). Bins with smaller dust species are more closely spaced than large-particle bins. }
\label{tab:DustSize}
\centering
\begin{threeparttable}
\begin{tabular}{l c c}    
\hline\hline
2 Fluids		& 4 Fluids		& 8 Fluids \\
\hline
			& 			& 7.23 nm \\
8.21 nm		& 7.89 nm		& 20.4 nm \\ 
			& 			& 39.4 nm \\
			& 44.2 nm		& 64.3 nm \\
			&			& 94.9 nm \\
117 nm		& 105 nm		& 131 nm \\
			& 			& 173 nm \\
			& 189 nm		& 221 nm \\
\hline        
\end{tabular}
     \vspace*{0.2cm}
  \end{threeparttable}
\end{table}

While we can conclude that the overall physics is captured in all these simulations, we note that minor small-scale differences are observed in the dust distribution. While the growth rate of the linear phase is comparable for all three simulations, differences in evolution during the non-linear phase are also observed between the simulation with two species on the one hand and the simulations with four and eight species on the other hand. Choosing for a setup with four dust species is therefore a good choice with the treatment that we described in section \ref{DusFlu}.

\end{document}